\documentclass[12pt]{elsarticle} 
\usepackage{setspace}
\usepackage[colorlinks=true, allcolors=blue]{hyperref} 
\usepackage[english]{babel}
\usepackage{physics}
\usepackage{tabularx}
\usepackage{amsfonts}
\usepackage{color} 
\usepackage{graphicx, multicol,float} 
\usepackage{caption}
\usepackage{subcaption}
\usepackage{lscape}
\usepackage[english]{babel}	
\usepackage[utf8]{inputenc}
\usepackage{physics}
\usepackage[normalem]{ulem}
\biboptions{sort&compress}
\usepackage[colorlinks=true, allcolors=blue]{hyperref}

\usepackage{multicol}
\usepackage{geometry}
\journal{Arxiv}
\begin{document}

\title{Topological changes and deformation mechanisms of nanoporous Ta under compression}

\author[fcen,ifeg,famaf]{N. Vazquez von Bibow}
\author[conicet]{E.N. Millán} 
\author[conicet,imdea]{C.J. Ruestes} 

\address[fcen]{Facultad de Ciencias
  Exactas y Naturales, Universidad Nacional de Cuyo, Mendoza 5500,
  Argentina}

\address[ifeg]{Instituto de Física Enrique Gaviola, Universidad Nacional de Córdoba UNC-CONICET, Facultad de Matemática, Astronomía, Física y Computación, Av. Medina Allende s/n, Ciudad Universitaria, 5000, Córdoba, Argentina.}
\address[famaf]{Facultad de Matemática, Astronomía, Física y Computación, 5000, Córdoba, Argentina.}
\address[conicet]{Instituto Interdisciplinario de Ciencias Básicas (ICB), Universidad Nacional de Cuyo UNCuyo-CONICET, Facultad de Ciencias Exactas y Naturales, Padre Contreras 1300, 5500, Mendoza, Argentina}
\address[imdea]{IMDEA Materials Institute, C/Eric Kandel 2, 28906, Getafe, Madrid, Spain}

\date{\today}

\begin{abstract}
While the mechanical behavior of noble nanoporous metals has been the subject of numerous studies, less is known about their recently developed refractory-based counterparts. Here we report on the mechanical properties, deformation mechanisms and topological changes of nanoporous tantalum, a prototypical refractory metal, by means of atomistic simulations of compression tests. An open-source multi-cpu and gpu-capable software is presented and used for the generation of computational samples. The stress strain curves show a non-linear elastic response, with early yielding. The plastic regime is first characterized by a linear hardening followed by an exponential hardening at large strains, associated with a high degree of densification. Plasticity is dominated by dislocation activity, with twinning and vacancy formation appearing as complementary deformation mechanisms. In order to study the mechanical response from a topological perspective, we track the evolution of the genus throughout the tests, finding direct correlations with each regime of the stress strain curves. 
The results are in agreement with previous studies of plasticity in nanoporous metals and highlight the importance of using topological metrics, for gaining insights into complex aspects of the deformation of nanoporous metals.
\end{abstract}
\maketitle

\section{Introduction}

Nanoporous (np) metals offer a vast range of applications. This includes unique performance for catalytic and optical applications due to its high surface-to-volume ratio \cite{ding2004metallic,wittstock2023nanoporous}. In addition, their development opened the door for the fabrication of surface-chemistry-powered actuators and sensors, among other cutting-edge technology \cite{biener2009surface,detsi2016metallic}. The range of potential applications extends to materials for advanced fission and fusion reactors \cite{bringa2011nanoporous,beyerlein2013radiation,briot2019situ,lionello2021mechanical}. Studies on the mechanical properties of nanoporous metals typically deal with noble metals, most notably nanoporous gold (np-Au) \cite{biener2006size,hodge2007scaling,feng2009surface,sun2013mechanical,farkas2013mechanical,briot2014mechanical,
luhrs2016elastic,ruestes2016hardening,jiao2017deformation,badwe2017mechanical,luhrs2017plastic,burckert2017uniaxial,pia2018nanoporous,
beets2019deformation,wilkerson2019anomalous,jin2018mechanical}. Despite the remarkable properties of np-Au, its thermal instability \cite{biener2011ald} and propensity to coarsening \cite{li2019topology} limits practical applications. 

The development of modern dealloying techniques opened the door for the fabrication of new non-noble nanoporous metals, such those made of refractory metals. Important examples from a technological application perspective include np-W, np-Ta, np-Mo and np-Nb \cite{mccue2015frontiers,geslin2015topology,mccue2016kinetics,hou2018nanoporous,kosmidou2019vacuum,gaskey2019self,chuang2022powder}. Despite their potential, experiments on mechanical characterization of such materials is generally lacking, with a notable exception after the work of Zhao et al. \cite{zhao2020tailoring,zhao2021open}, who used nanoindentation testing to probe the mechanical properties of np-W with relatively high solid fraction  (above 0.484). Their studies show that np-W is a high performance material, with a high hardness (several GPa), highlighting its potential applications in harsh environments, such as nuclear facilities. 


The mechanical properties of macroporous foams are typically dictated by the relative density and characteristics of its base material. In contrast, the mechanical properties of nanoporous metals are not only dictated by the former, but also by their ligament size \cite{hodge2007scaling}, surface effects \cite{detsi2011specific} and network connectivity \cite{liu2016interpreting} among other topological aspects \cite{mangipudi2016topology,lilleodden2018topological}. 
 
Molecular dynamics (MD) simulations, despite timescale and lengthscale limitations, have proven to be a valuable tool for studying the mechanical response of nanoporous metals. Two notable contributions serve as examples. The first one is that of Farkas et al. \cite{farkas2013mechanical}, who first reported on tension-compression asymmetry for nanoporous gold, confirmed experimentally several years later \cite{luhrs2017plastic}. The second one is that of Bringa et al. \cite{bringa2011nanoporous}, who showed that nanoporous metals can display radiation tolerance depending on irradiation conditions and ligament size, opening a new line of application for these materials \cite{beyerlein2013radiation}. In addition, the MD technique has been used to provide important insights into a variety of effects. A non-exhaustive list includes ligament size effects \cite{li2018mechanical,saffarini2021ligament}, strain-rate effects \cite{yildiz2020strain,voyiadjis2021characterization}, grain size effects \cite{li2018mechanical,li2020nanoindentation}, temperature effects \cite{saffarini2021temperature}, anomalous compliance and early yielding \cite{ngo2015anomalous}, densification and hardening \cite{ruestes2016hardening,saffarini2021ligament} and load-bearing network and topological effects \cite{beets2018mechanical,mathesan2020size,mathesan2021yielding}. The technique has even been used to derive and inform scaling laws \cite{sun2013mechanical,beets2019deformation,saffarini2021scaling} and also to provide fundamental explanations on np-Au fracture \cite{beets2020fracture}. More recently, it has provided valuable insights into the mechanical response of graded nanoporous structures \cite{he2022mechanical} and hierarchical nanoporous structures \cite{shi2021scaling}. 

The knowledge on the mechanical properties of nanoporous refractory metals is still in its infancy compared to that of noble metals, for which much has been investigated using nanoporous gold \cite{jin2018mechanical}. Differences in its base material, slip systems, dislocation interaction and storage are expected to play an important role in the mechanical response. However, these aspects have not been deeply investigated so far and are brought into focus in this study. In fact, dislocation accumulation in nanoporous gold has been linked to its hardening behavior under compression, together with uniform densification. In can be expected that if densification takes place, the connectivity of the foam would change and if so, the strain hardening behavior of nanoporous foams could be correlated to changes in its topology. This leads to the conjecture that alterations in the foam's topology, stemming from the occurrence of densification, may significantly influence the strain hardening behavior of nanoporous foams. Here 
we test this hypothesis employing surface reconstruction techniques and common topological metrics.









In this work, we explore the mechanical response, deformation mechanism and topology evolution during compression of nanoporous tantalum, a prototypical refractory metal. An open-source multi-cpu and gpu-capable software is presented and used for the generation of computational samples. We perform MD-simulated compression tests up to large strains, in excess of 60 percent, fully exploring the elastic regime, linear hardening regime and exponential densification regime. In addition to the inspection of the deformation mechanisms, we present an analysis of the evolution of topology during deformation.
Finally, we correlate the different hardening stages with notorious changes in the topology of the foams.

\section{Computational modeling}

\subsection{Code for generation of nanoporous samples}
\label{sample}

Nanoporous tantalum produced by liquid-metal dealloying can be classified as an open-cell foam with nanoscale features \cite{mccue2016kinetics}. It shows striking similarities with nanoporous gold produced by dealloying \cite{ding2004metallic,hodge2006characterization,beyerlein2013radiation}, resembling those structures associated with spinodal decomposition \cite{newman1999alloy}. Initial attempts to study nanoporous gold with MD simulations relied on sample generation by means of phase-field methods \cite{crowson2007geometric,crowson2009mechanical}. More recently, the development of levelled-wave methods based on
Cahn’s equation \cite{soyarslan20183d,liu2019efficient}, and even fully molecular dynamics methods
based on Ag-Au demixing \cite{guillotte2019fully} resulted in a significant increase of studies of np metals using MD simulations. In this work, we employ an in-house implementation of the levelled-wave method presented by Soyarslan et al. \cite{soyarslan20183d}. This method has been successfully applied in a number of studies \cite{mathesan2021yielding,he2022mechanical,mathesan2020size,shi2021scaling,li2019topology}. For completeness, the method is briefly described below.

In 1965, Cahn presented the theory of phase separation from a single phase fluid by a spinodal mechanism. He showed that the predicted
structure may be described in terms of the superposition of an infinite number of sinusoidal waves of a fixed
wavelength, but random in amplitude, orientation, and phase \cite{cahn1965phase}. More recently, Soyarslan et al. \cite{soyarslan20183d} explore a modern implementation of this random field $f$:

\begin{equation}
    f(\vb{x}) = \sqrt{\frac{2}{N}}\sum_{i=1}^N \cos(\vb{q_i}\vdot\vb{x}+\varphi_i).
    \label{eq:Soyarslan_RandomField}
\end{equation}
where $\vb{x}$ is the position vector, $N$ is the number of waves considered on the truncated series, while $\vb{q_i}$ and $\varphi_i$ denote the direction and phase of the wave $i$, respectively. The wave number must be set to a constant value (i.e. $\abs{\vb{q_i}} = q_0$), with directions uniformly distributed over a $4\pi$ solid angle, and phases uniformly distributed between $0$ and $2\pi$. Under this constraints, $f(\vb{x})$ is a random gaussian field with $\expval{f}=0$ and $\expval{f^2}=1$.
Provided N is sufficiently large, the value $f(\vb{x)}$ for a given position  $\vb{x}$ follows a Gaussian distribution: 

\begin{equation}
    P(f) = \frac{1}{\sqrt{2\pi}}e^{-f^2/2},
    \label{eq:Soyarslan_fGauss}
\end{equation}
and the mean wave direction  
\begin{equation}
    \vb{\nu} = \frac{1}{N} \sum_{i=1}^N \vb{n_i}
\end{equation}
tends to zero as consequence of the central limit theorem. Therefore, $f$ is an isotropic function. 

For equation \eqref{eq:Soyarslan_RandomField}, the different phases of the system are obtained by means of a cutoff value $\xi$: 

\begin{equation}
\begin{split}
    &\vb{x}\in\mathcal{B}\qquad\text{if}\,f(\vb{x})<\xi,\\
    &\vb{x}\in\partial\mathcal{B}\;\,\quad\text{if}\,f(\vb{x})=\xi,\\
    &\vb{x}\in\mathcal{P}\qquad\text{if}\,f(\vb{x})>\xi.
\end{split}
\label{eq:Soyarslan_CutValuexi}
\end{equation}
Here, $\mathcal{B}$ is the solid phase, $\mathcal{P}$ is the porous phase and $\partial\mathcal{B}$ is the boundary between the two phases.
$\xi$ can be correlated to the solid fraction, defined as $\phi_\mathcal{B}:=\abs{\mathcal{B}}/\abs{\mathcal{V}}$ (the ratio of the volume of the solid phase to the total volume of the sample), thanks to the properties of the random Gaussian field:

\begin{equation}
    \xi(\phi_\mathcal{B}) = \sqrt{2}\operatorname{erf}^{-1}(2\phi_\mathcal{B}-1),
    \label{eq:Soyarslan_xiPhi_b}
\end{equation}

The structures generated using eqns.  \eqref{eq:Soyarslan_RandomField} and \eqref{eq:Soyarslan_CutValuexi} are, in general, non periodic. Periodicity can be obtained by choosing a finite number of waves with an integer wave number in all directions, with constant modulus.  

Starting from \eqref{eq:Soyarslan_RandomField}, with $\vb{e_1,e_2,e_3}$  unit vectors of an orthonormal base in real space, $\vb{q_i}$ must be of the form

\begin{equation}
    \vb{q} = \frac{2\pi}{A}(h,k,l),
    \label{eq:Soyarslan_q}
\end{equation}
where $h,k,l\in\mathbb{Z}$ correspond to Miller indexes and $A$ is a constant. It can be verified that $f(\vb{x}) = f(\vb{x}+n_1A\vb{e_1}+n_2A\vb{e_2}+n_3A\vb{e_3}),$ with $n_1,n_2$ y $n_3$ arbitrary integers. This implies that $f$ is periodic with magnitude $A$ with respect to the base. Moreover, if \eqref{eq:Soyarslan_RandomField} is taken with identical $\phi_i$, then $f$  is also invariant with respect to an axis interchange. So the resulting $f$ has cubic symmetry with grid parameter $A$. 
Based on the aforementioned considerations, periodic stochastic structures can be built by restricting the sum on eqn. \eqref{eq:Soyarslan_RandomField} to a set of  $\vb{q_i}$ with a given value and constant $H$, so that
\begin{equation}
H = \sqrt{h^2+k^2+l^2},
\label{eq:H_Soyarslan}
\end{equation}
with $\abs{\vb{q}}=q_0=2\pi H/A$. 

For structures with equal fraction of the two phases, symmetry suggests that the average ligament size is equal to the average pore size: 

\begin{equation}
    L = 1.23 \frac{\pi}{q_0}
    \label{eq:Soyarslan_L1}
\end{equation}

For nanoporous structures with solid fraction different than 0.5, the average ligament size of the solid phase is given by: 
\begin{equation}
    L = \frac{A}{H}[0.53\phi_B+0.41].
    \label{eq:Soyarslan_L2}
\end{equation}
For more details on the method, the reader is referred to \cite{cahn1965phase,soyarslan20183d}.

While the implementation of the method is straightforward by means of nested-loop programming, such approach is not practical given memory and wall-time limitations. Therefore, we opted for a more elaborated approach. We define a rank-3 tensor 

    \begin{equation}
        T_{ijk} := (\vb{q\vdot x})_{ij} + \phi_k,
    \end{equation}

for all points $\vb{x}$ of the domain and for all $\vb{q}$ wave vectors on the set of directions. The implementation makes use of  \textit{NumPy} \footnote{\url{https://numpy.org/}} . In order to avoid memory limitations, the $T_{ijk}$ tensor was decomposed into smaller tensors. Such segmentation allows for multi-core parallelization using \textit{joblib} \footnote{\url{https://joblib.readthedocs.io/en/stable/}}   and \textit{loky}\footnote{\url{https://github.com/joblib/loky}} libraries. Finally, the code was also implemented on GPUs using  \textit{Tensorflow 2} framework \footnote{\url{https://www.tensorflow.org/}}. Benchmarking of the code is presented in section \ref{Appendix}. The code is provided open-source \footnote{\url{https://github.com/nicolasvazquez95/Nanofoams_Creator}}. 

\subsection{Nanoporous samples and their relaxation}

The method explained in the previous section was used to produce stochastic bicontinuous structures of two phases, $\alpha$ and $\beta$. Equation \ref{eq:Soyarslan_CutValuexi} was sampled at the atomic locations of a [100]-oriented Ta single crystal and all atoms located in phase $\alpha$ -- as determined by the threshold parameter $\xi$ -- were removed to create pores, with the nanoscale porous structure resulting from the remaining phase $\beta$ atoms. The samples were generated using periodic boundary conditions, with a lattice parameter $a_0 = 0.3303 ~nm$, $N=20000$ waves, $H^2=161$ and a solid fraction $\phi_B \approx 0.3$. 
We generated three virtual np structures, each with a different ligament size, see Table \ref{tab:Muestras}.


\begin{table}
\begin{tabular}{lll|ll|ll|ll|}
\cline{4-9}
\multicolumn{3}{l|}{}                                                                                                                                                 & \multicolumn{2}{l|}{As-generated}                                              & \multicolumn{2}{l|}{\begin{tabular}[c]{@{}l@{}}After \\ relaxation\end{tabular}} & \multicolumn{2}{l|}{\begin{tabular}[c]{@{}l@{}}Mechanical \\ properties\end{tabular}}                                              \\ \hline
\multicolumn{1}{|l|}{\begin{tabular}[c]{@{}l@{}}Simulation\\ domain\end{tabular}} & \multicolumn{1}{l|}{Atoms} & \begin{tabular}[c]{@{}l@{}}$L_{box}$\\  (nm)\end{tabular} & \multicolumn{1}{l|}{$\phi_B$} & \begin{tabular}[c]{@{}l@{}}L \\ (nm)\end{tabular} & \multicolumn{1}{l|}{$\phi_B$}  & \begin{tabular}[c]{@{}l@{}}L \\ (nm)\end{tabular}  & \multicolumn{1}{l|}{\begin{tabular}[c]{@{}l@{}}E \\ (GPa)\end{tabular}} & \begin{tabular}[c]{@{}l@{}}$\sigma_y$ \\ (MPa)\end{tabular} \\ \hline
\multicolumn{1}{|l|}{150 $a_0$}                                                      & \multicolumn{1}{l|}{2.3 10$^6$}      &  49.5                                                    & \multicolumn{1}{l|}{0.291} & 2.26                                              & \multicolumn{1}{l|}{0.294}       & 2.4                                                & \multicolumn{1}{l|}{1.80}                                               & 89.3                                                     \\ \hline
\multicolumn{1}{|l|}{200 $a_0$}                                                      & \multicolumn{1}{l|}{5.5 10$^6$}      &    66                                                  & \multicolumn{1}{l|}{0.301} & 3.1                                               & \multicolumn{1}{l|}{0.303}       & 3.1                                                & \multicolumn{1}{l|}{1.83}                                               & 91.0                                                     \\ \hline
\multicolumn{1}{|l|}{250 $a_0$}                                                      & \multicolumn{1}{l|}{10.3 10$^6$}      &    82.5                                                  & \multicolumn{1}{l|}{0.297} & 3.78                                              & \multicolumn{1}{l|}{0.298}       & 3.8                                                & \multicolumn{1}{l|}{2.18}                                               & 98.4                                                     \\ \hline
\end{tabular}
	\caption{Summary of sample characteristics and dimensions. These correspond to the as-generated samples (i.e. before relaxation). Relative densities were computed using the \textit{ConstructSurfaceMesh} algorithm implemented in  OVITO. Average ligament sizes $L$ were computed using FoamExplorer \cite{aparicio2020foamexplorer}.}
\label{tab:Muestras}
\end{table}

Prior to the simulation of mechanical tests, each foam was first minimized by the conjugate gradient method specifying a force tolerance of $10^{-8}$ eV/Angstrom and an energy tolerance of $10^{-6}$ eV. Then the structure was thermally relaxed at 300 K for 500 ps and zero pressure using a NPT ensemble. Periodic boundary conditions were imposed in all directions. The simulations were performed using LAMMPS \cite{plimpton1995fast} and the Ta atomic interactions were modeled by means of an extended Finnis-Sinclair potential (EFS) \cite{dai2006extended}. This potential successfully reproduces lattice constants, cohesive energies, elastic constants and vacancy formation energies of Ta \cite{dai2006extended}. The generalized stacking faults energy curves and generalized twinning fault energy curves had been thoroughly studied \cite{tang2011growth,tang2012growth,ruestes2014atomistic} and compared to other Ta potentials \cite{ruestes2014atomistic}. The potential has proven useful in describing dislocation-mediated plasticity in closed-cell nanoporous Ta under tension and compression \cite{tang2011growth,tang2012growth,ruestes2013atomistic,tramontina2014orientation}, as well as bulk Ta under spherical nanoindentation \cite{ruestes2014atomistic}.



\begin{figure}[H]
    \centering
    \includegraphics[width=\textwidth]{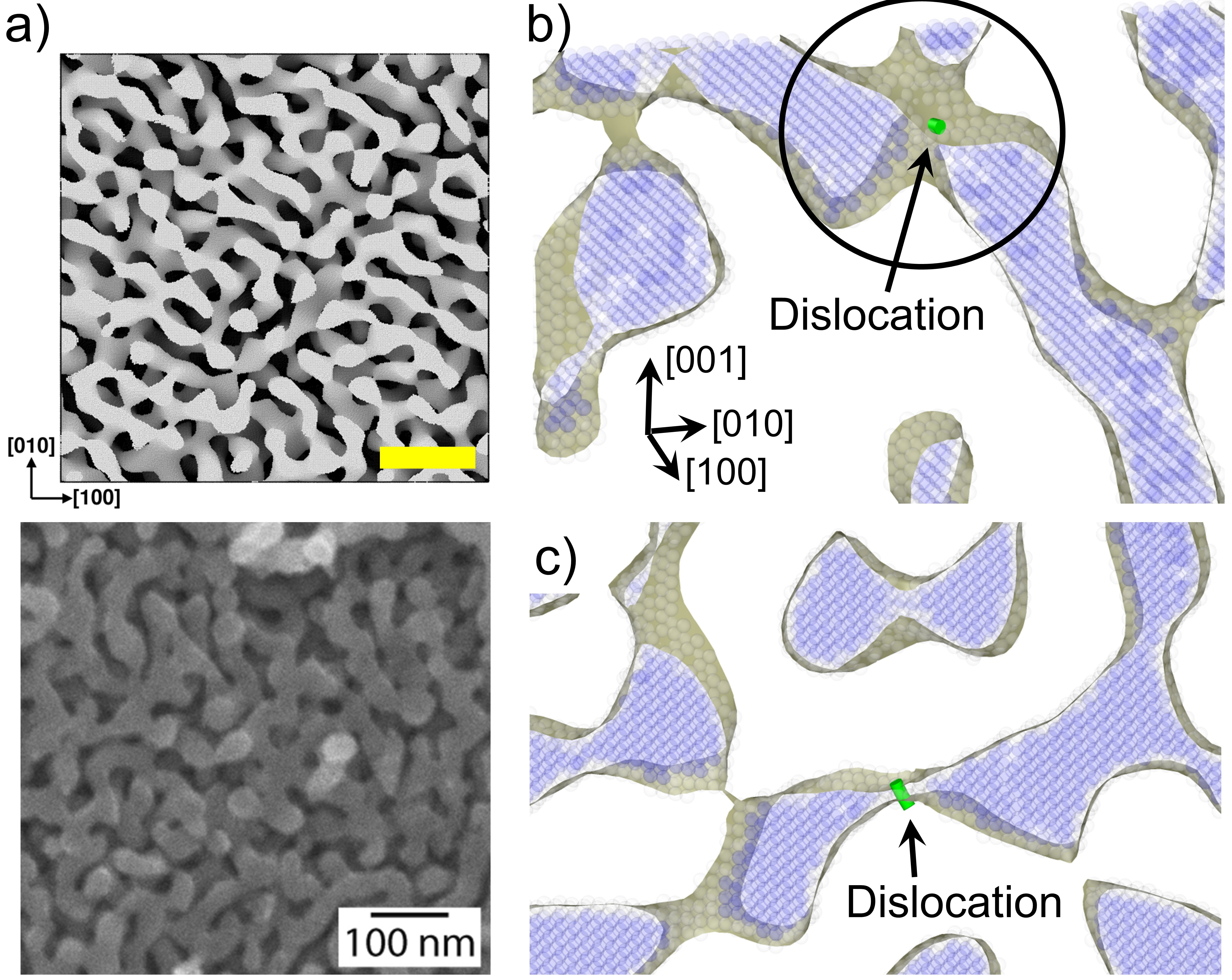}
    \caption{a) Plan view of the largest sample generated (Simulation domain 250 $a_0$, $L_{box}$ =82.5 nm, Scale bar 20 nm) compared to a nanoporous Ta sample produced by liquid metal dealloying (Reprinted from \cite{mccue2016kinetics}, with permission from Elsevier). b) Dislocation (green) nucleated during relaxation on a ligament junction (circle). c) Another example of dislocation generated during relaxation. }
    \label{fig:RelaxedSample}
\end{figure}

As a result of the minimization and relaxation strategy, a small amount of ligament coarsening and a reduction of the sample dimensions was verified, and the updated dimensions are presented on Table \ref{tab:Muestras}. These correspond to converged values, i.e. they remained unaltered during the last 200 ps of the relaxation stage. Relaxation of the initially rigid and defect-free porous crystal in the absence of mechanical loading led to samples with a number of dislocations (Fig. \ref{fig:RelaxedSample}). Similar defects had been reported after transmission electron micrographs of nanoporous gold produced by dealloying \cite{dou2011deformation}. 


As a result of the whole procedure, we were able to generate open-cell np-Ta atomistic samples with pre-existing defects, that closely resembles np-Ta foams produced by liquid-metal dealloying \cite{mccue2016kinetics}, combining nanoscale ligaments and pores. According to \cite{soyarslan20183d}, the generated microstructures can be considered as representative volume elements (RVEs) given the parameters used on their generation.

\subsection{Simulation details}
\label{simdet}

Uniaxial compression tests were simulated with the loading axis aligned to the z-axis. A thermostat and barostat were used to maintain lateral pressure (x and y directions) at approximately 0 Pa and to keep temperature constant at 300K. Mimicking strain-controlled compressive mechanical deformation tests, each sample was uniaxially compressed along the [001]-direction by re-scaling its z-dimension at a strain rate of 1 × 10$^{8}$ s$^{-1}$ with a 2 fs timestep.  
Stresses are presented in the form of von Mises stress. LAMMPS output of the stress tensor of the simulation box is $\sigma_{ij}$ from which the scalar quantity $\sigma_{VM}$ can be computed as:

\begin{equation}
    \sigma_{VM} = \sqrt{\sigma_{xx}^2+\sigma_{yy}^2+\sigma_{zz}^2 - (\sigma_{xx}\sigma_{yy} + \sigma_{xx}\sigma_{zz} + \sigma_{yy}\sigma_{zz}) + 3(\sigma_{xy} + \sigma_{xz} + \sigma_{yz})}
    \label{eq:sigma_VM}
\end{equation}

Post-processing was performed using OVITO \cite{stukowski2010visualization} and the Crystal Analysis Tool \cite{stukowski2012automated}, thus allowing for detailed defect identification and tracking.

\section{Results}

\subsection{Stress - strain curves}

\begin{figure}[h]
    \centering
    \includegraphics[width=0.8\textwidth]{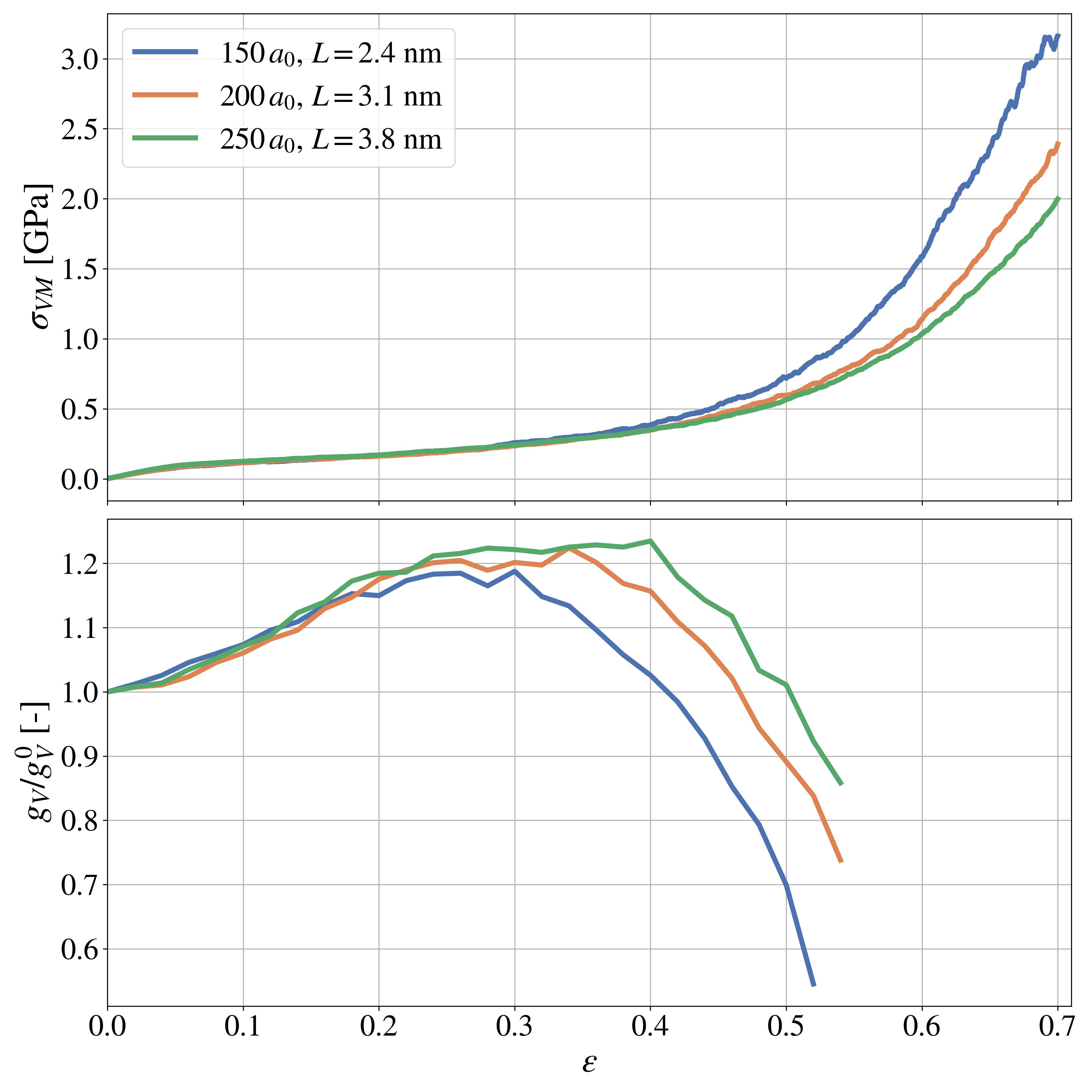}
    \caption{\textit{Top} von Mises stress - strain plot for the three samples studied, according to their average ligament size. \textit{Bottom} Dimensionless genus per volume ratio ( $g_V/g_{V}^0$) for the three samples studied.}
    \label{fig:stress_genus}
\end{figure}

\begin{figure}[h]
    \centering
    \includegraphics[width=0.8\textwidth]{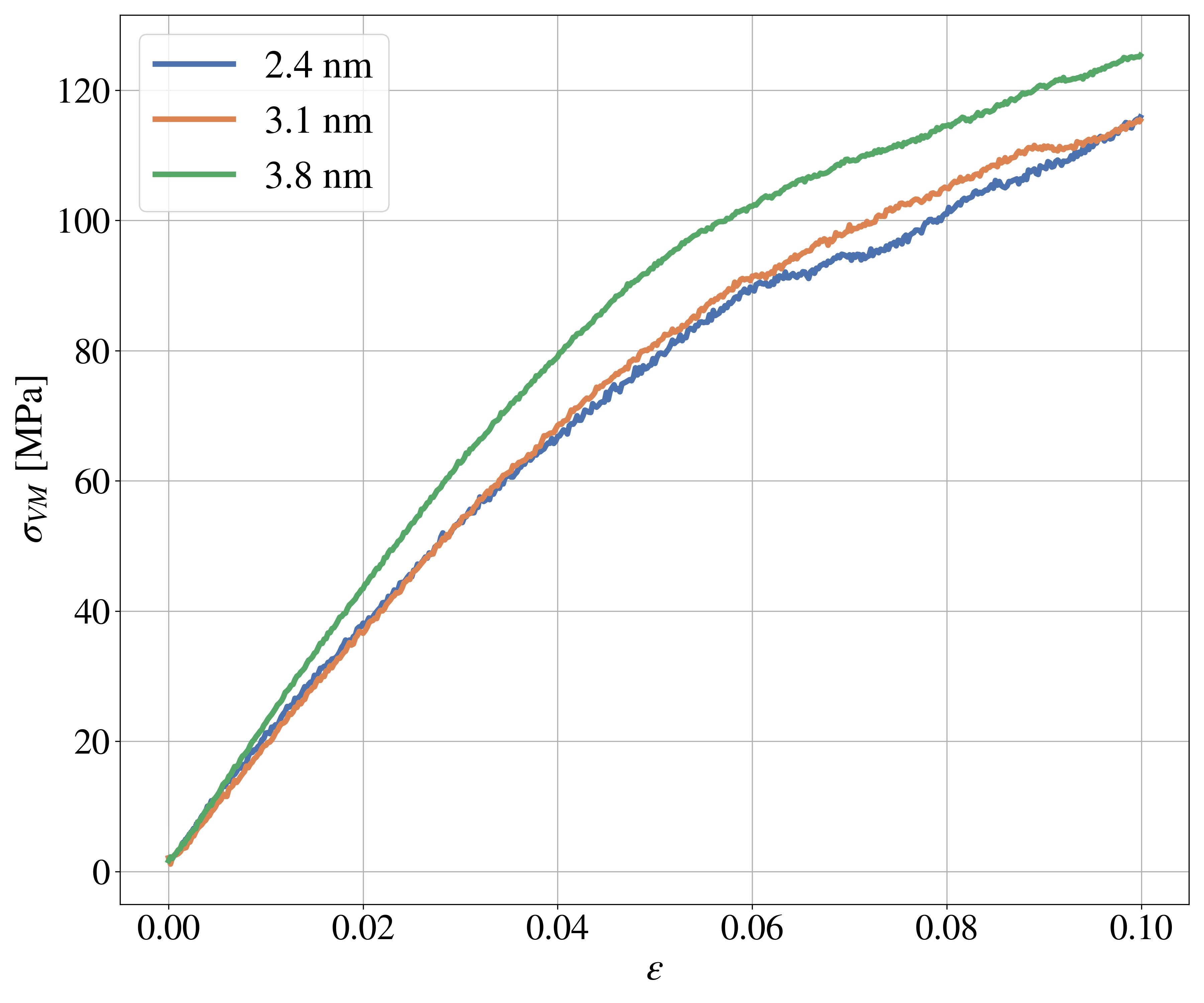}
    \caption{Early stage of the von Mises stress - strain plot for the three samples studied, according to their average ligament size.}
    \label{fig:elastic}
\end{figure}

Figure \ref{fig:stress_genus} (top) presents the stress-strain curves for the three samples studied. Figure \ref{fig:elastic} presents an extract of the stress- strain curve focusing on the elastic regime and the early stages of plasticity. Noticeable features included a non-linear elastic regime (\ref{fig:elastic}), as well as early-yielding. Both behaviors had been previously reported for nanoporous gold \cite{ngo2015anomalous,jin2018mechanical}.

Table \ref{tab:Muestras} presents the elastic modulus $E$ and yield stress $\sigma_Y$ obtained for our samples. The former was obtained from the stress strain response during the first 0.01 strain. The latter was taken as the intersection of the stress strain curve with a parallel line offset by 0.01 strain, its slope being $E$. Both elastic modulus and yield strength decrease with ligament size, in a \textit{smaller is weaker} fashion. This is attributed to the ligament size range explored (below 5 nm), akin to an inverse Hall-Petch effect.

All simulations present a linear hardening regime for strains below 0.4. 
While hardening is typically absent for macroporous foams \cite{gibson1997cellular}, it is not new for nanoporous metals. Linear hardening in nanoporous Au under compression is known to be ligament size-dependent. This was first reported in experiments \cite{jin2009deforming}, but can also be found in molecular dynamics simulation literature \cite{saffarini2021ligament}. In our case, all three samples display a similar linear hardening stage, almost overlapping, most likely due to ligament sizes within the same order of magnitude, unlike in \cite{jin2009deforming,saffarini2021ligament}. Under compression, nanoporous metals typically densify uniformly, leading to a continuous hardening response \cite{jin2009deforming}. As a consequence, the onset of densification regime is ill-defined for nanoporous metals, and it is known to vary with ligament size. That is clearly seen in our curves. Samples with smaller ligament size densify earlier than those with larger ligament size, in agreement with experimental findings for np-Au \cite{jin2018mechanical}.
Interestingly, the exponential hardening regime shows a noticeable ligament size dependence, and will be further discussed in Sec. \ref{TopoSection}. 

\subsection{Deformation mechanisms}
\label{sec:defmech}
\begin{figure}[h]
    \centering
    \includegraphics[width=0.8\textwidth]{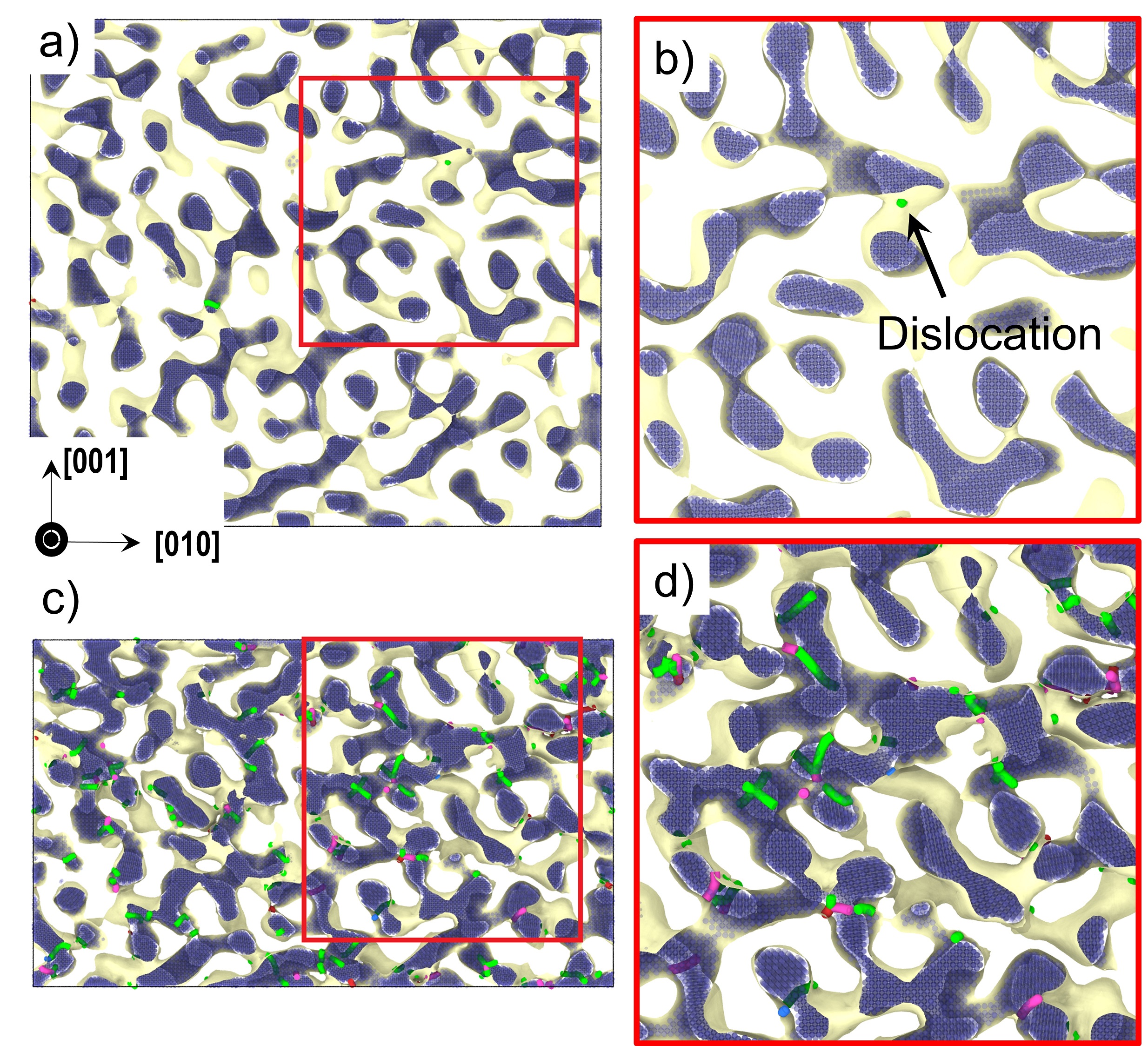}
    \caption{Dislocation activity. (a,b) At a strain of 0.1, few dislocations are noticeable on some ligaments. Most of the generated dislocations have reached nearby surfaces, effectively annihilating. (c,d) At a strain of 0.4, a large number of dislocations are seen, mainly on ligament junctions. These are glissile dislocations (green - $\Vec{b} = \frac{a_0}{2} \langle 111 \rangle$) and sessile dislocations (pink - $\Vec{b} = a_0 \langle 100 \rangle$). Note that the densification has started.}
    \label{fig:defmech}
\end{figure}

Initially and up to moderate strains, our simulations reveal that nanoporous Ta deforms by ligament bending with plastic collapse at the ligament nodes or close to them (Figure \ref{fig:defmech}). Plasticity mainly proceeds by dislocation activity . Nucleation takes place at free surfaces in the form of glissile dislocations (Figure \ref{fig:defmech}.a,b). They propagate across the ligaments and annihilate at nearby surfaces. These corresponds to dislocations with Burgers vector $\Vec{b} = \frac{a_0}{2} \langle 111 \rangle$. Despite the abundance of free surfaces that act as dislocation sinks, deformation proceeds in a dislocation accumulation scenario, in agreement with experimental and computational studies on nanoporous gold \cite{jin2009deforming,ruestes2016hardening}. It must be noted that under creep conditions, np-Au MD studies suggest that nanoporous metals can deform by displacive-diffusion mechanisms \cite{mathesan2023displacive}. We found no indications of such mechanism, most probably due to the high strain-rate conditions and short timescales explored.

In addition to surface exhaustion mechanisms, a fraction of dislocations get retained inside the ligaments. Such retention is favored at ligament junctions, where dislocations have enough space for reaction with other dislocations, promoting junction formation as well as the generation of sessile dislocations  ($\Vec{b} = a_0 \langle 100 \rangle$) by the following reaction,

\begin{equation}
    \frac{1}{2} \langle 1\bar{1}\bar{1} \rangle + \frac{1}{2} \langle \bar{1}\bar{1}1 \rangle \to \langle 0\bar{1}0 \rangle 
    \label{eq:disloc_reac_1}
\end{equation}

\begin{figure}[H]
    \centering
    \includegraphics[width=\textwidth]{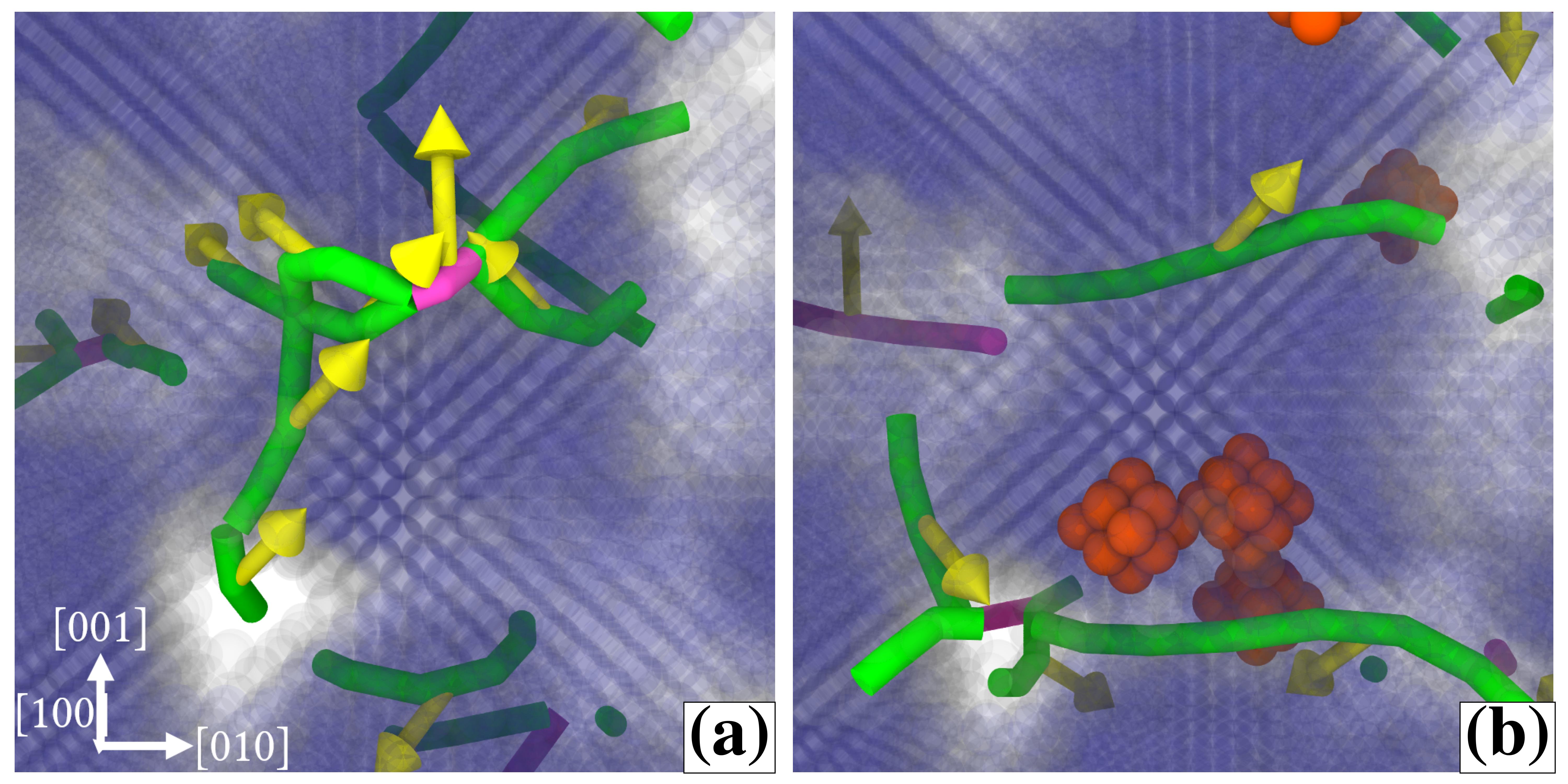}
    \caption{Dislocation junctions in a region with high dislocation density. Reaction of dislocations of glissile dislocations ($\Vec{b} = \frac{a_0}{2}\langle 111 \rangle$, green) lead to the generation of sessile dislocations ($\Vec{b} = a_0 \langle 100 \rangle$, pink). Dislocation reactions also promote vacancy generation. Arrows indicate Burgers vector. Reconstructed surfaces of the foam are not displayed to favor the visualization of dislolcations. Red clusters correspond to vacancies.}
    \label{fig:DislocReac_Vacancia}
\end{figure}

Prominent examples of the aforementioned mechanism can be seen on Figure \ref{fig:DislocReac_Vacancia}.a. It is also shown that vacancy formation also takes place due to the reaction of dislocations (Figure \ref{fig:DislocReac_Vacancia}.b). It is likely that self-interstitials get generated too. However, we failed to identify them. In MD simulations, self-interstitials are generally identified by means of a Wigner-Seitz analysis \cite{soltani2018mechanism} comparing a deformed configuration against an undeformed one. Such a comparison was not possible here due to the significant amount of deformation. It must also be noted that point defect generation might be influenced by strain-rate effects \cite{dupraz2018dislocation}.

\begin{figure}[h]
    \centering
    \includegraphics[width=0.8\textwidth]{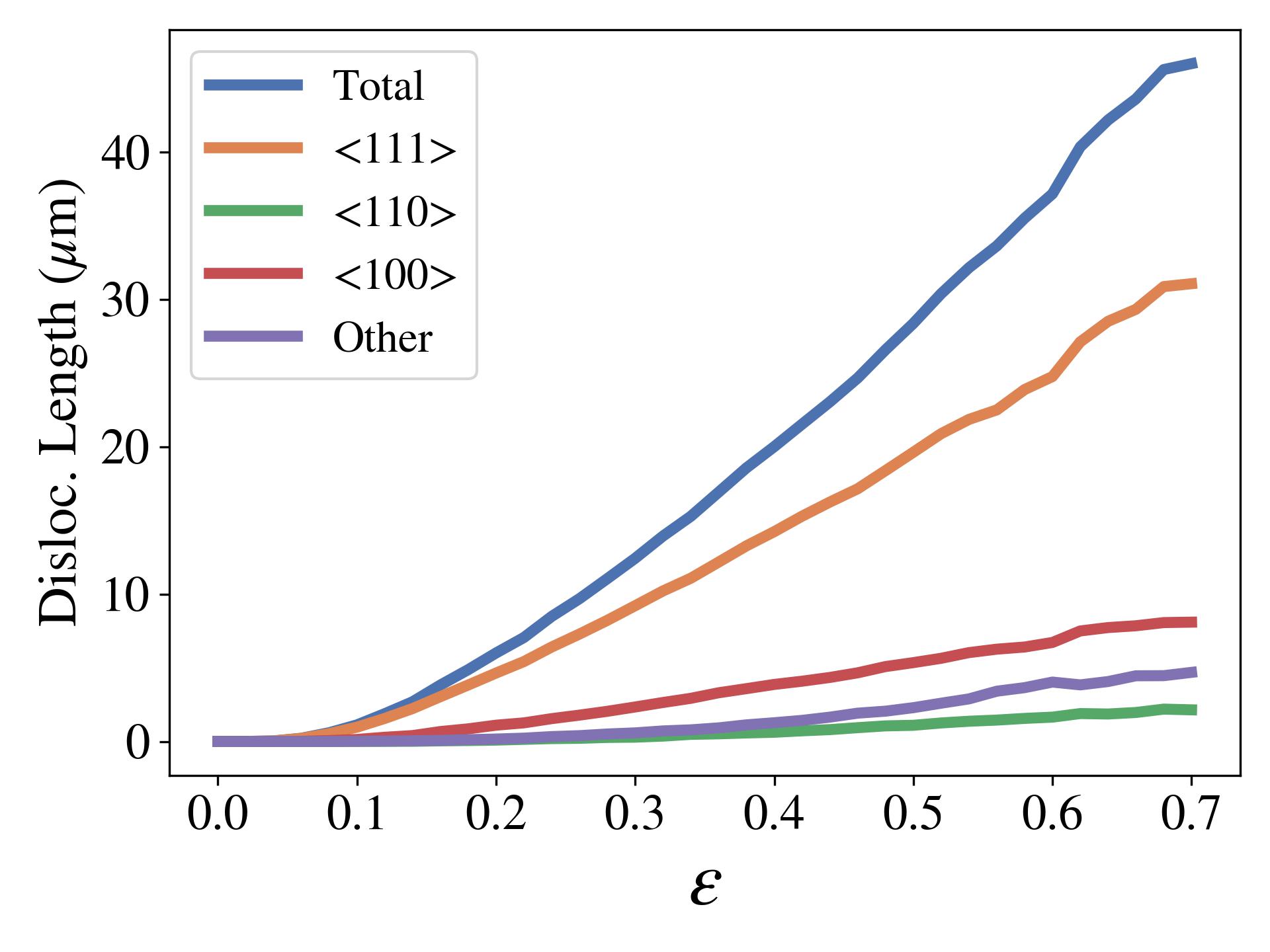}
    \caption{Evolution of total dislocation length for our largest sample tested (250 $a_0$, $L=3.8~nm$). Contributions to the total length include $<111>-type$ glissile dislocations, $<110>-type$ and $<100>-type$ sessile dislocations and 'Other'-type dislocations, as described in Sec. \ref{sec:defmech}.}
    \label{fig:disl_quantities}
\end{figure}

Figure \ref{fig:disl_quantities} presents an evolution of the total dislocation length for our largest sample. In addition, the total length is discriminated considering the burgers vector of the dislocations. It can be seen that $\langle 111 \rangle$ glissile dislocations dominate, followed by $\langle 100 \rangle$ and $\langle 110 \rangle$ sessile dislocations. Dislocations with different burgers vector are also identified by the algorithm (type 'Other'). However, detailed inspection of them reveals that the mostly have a burgers vector proportional to one of the previous three categories, while others are assigned a burgers vector not consistent with dislocations in bcc metals pointing to possible artifacts in the identification of dislocations in heavily strained configurations. Similar trends were observed for all the samples simulated here.

\begin{figure}[H]
    \centering
    \includegraphics[width=.95\textwidth]{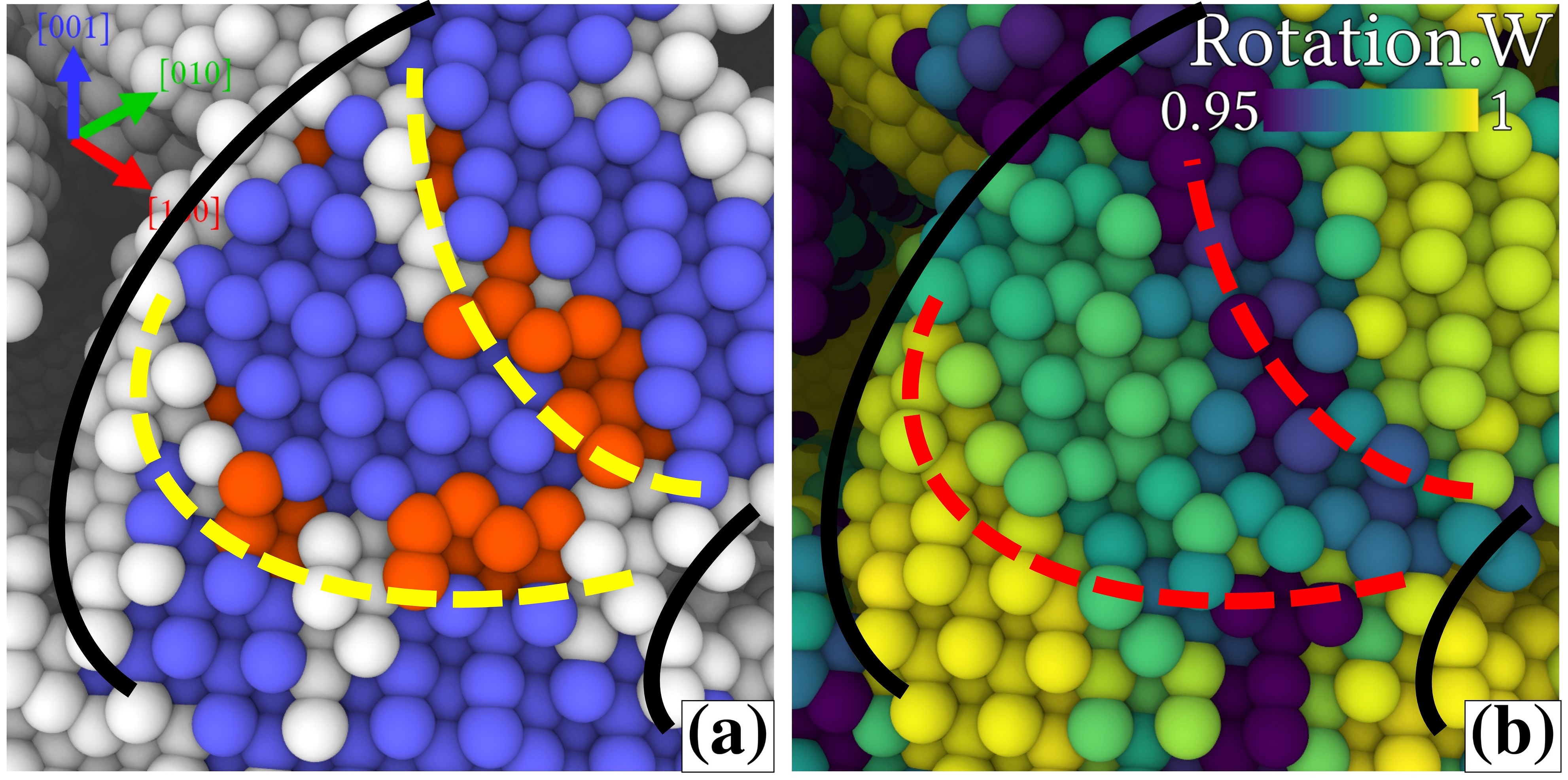}
    \caption{Example of a twin generated during deformation. \textbf{(a)} Atoms colored according to local structural environment: blue (bcc), red (twin boundaries), white (\textit{other}). \textbf{(b)} Atoms colored according to degree of rotation. Continuous lines indicate ligament surfaces. Dashed lines indicate twin boundaries. Note the rotation of the atoms between them.}
    \label{fig:Twins_Snapshot}
\end{figure}

A small fraction of twinning was also found, see Figure \ref{fig:Twins_Snapshot}. Tantalum is known to deform by twinning under high strain-rate compression \cite{murr1997shock}. High strain-rate conditions are inherent to MD simulations when exploring uniaxial compression of metals and therefore, this finding is not surprising. Experimental studies of nanoporous Ta at low strain rates are necessary to determine if twinning takes place in absence of high strain rates. 

\subsection{Topological evolution during deformation} \label{TopoSection}


A positive aspect of atomistic simulations is that they allow for detailed tracking of atomic positions and microstructures. In this way, and making use of surface reconstruction algorithms \cite{stukowski2014computational}, it is posible to track the evolution of the topology throughout the simulated compression test. 

At all stages of deformation, the surface of the sample can be determined by surface reconstruction techniques. Here we used the Gaussian density method, as implemented in OVITO \cite{krone2012fast}. Detailed information on the reconstructed surface mesh is readily available. This includes: number of vertices (V), number of edges (E) and number of faces (F) forming the reconstructed closed surface. 
This allows computing the Euler characteristic ($\chi$) as

\begin{equation}
    \chi = V-E+F
\end{equation}

For a closed surface, the Euler characteristic is equivalent to the genus ($g$) through:

\begin{equation}
    g = 1 - \frac{\chi}{2}
\end{equation}

This is a metric of the surface connectivity; it can be
interpreted as the number of continuous holes, or handles, on the resulting manifold made by the closed surface.

The determination of $\chi$ and $g$ after the reconstructed surface of the nanoporous sample is a valid procedure used both in experimental approaches, through FIB-nanotomography methods \cite{mangipudi2016fib}, and in molecular dynamics studies \cite{guillotte2019fully}.

In order to take into account volume changes, the genus is often normalized by the total volume of the foam and scaled using a characteristic length to produce a non-dimensional quantity \cite{soyarslan20183d}. We normalized the genus by the sample volume $g_V = g/V$. Then, by taking the ratio of genus
$g_V/g_{V}^0$, $g_{V}^0$ being the genus per volume in the absence of strain, we obtain the results in a non-dimensional form.



Figure \ref{fig:stress_genus} presents a correlation of the stress strain curves (a), with the $g_V/g_{V}^0$  ratio (b).   
Interestingly, the evolution of the genus ratio for the three samples also overlaps for the same strain regime that the stress strain curves do. Most importantly, the genus increases during the linear hardening regime and that increase starts right after the onset of plasticity, indicating that continuous densification is present through out this regime.

As the deformation of the foam enters the plastic regime, the foam experiences a compaction where more and more ligaments become into contact. As more ligaments become into contact, then one would expect that more cuttings would be needed to produce a disconected manifold. This can be seen on Figure \ref{fig:defmech} by comparing panels b) and d). Such behavior holds until strains of 0.4 (small sample - $L=2.4~nm$) to 0.5 (large sample - $L=3.8~nm$) are reached. 

\begin{figure}[H]
    \centering
    \includegraphics[width=.95\textwidth]{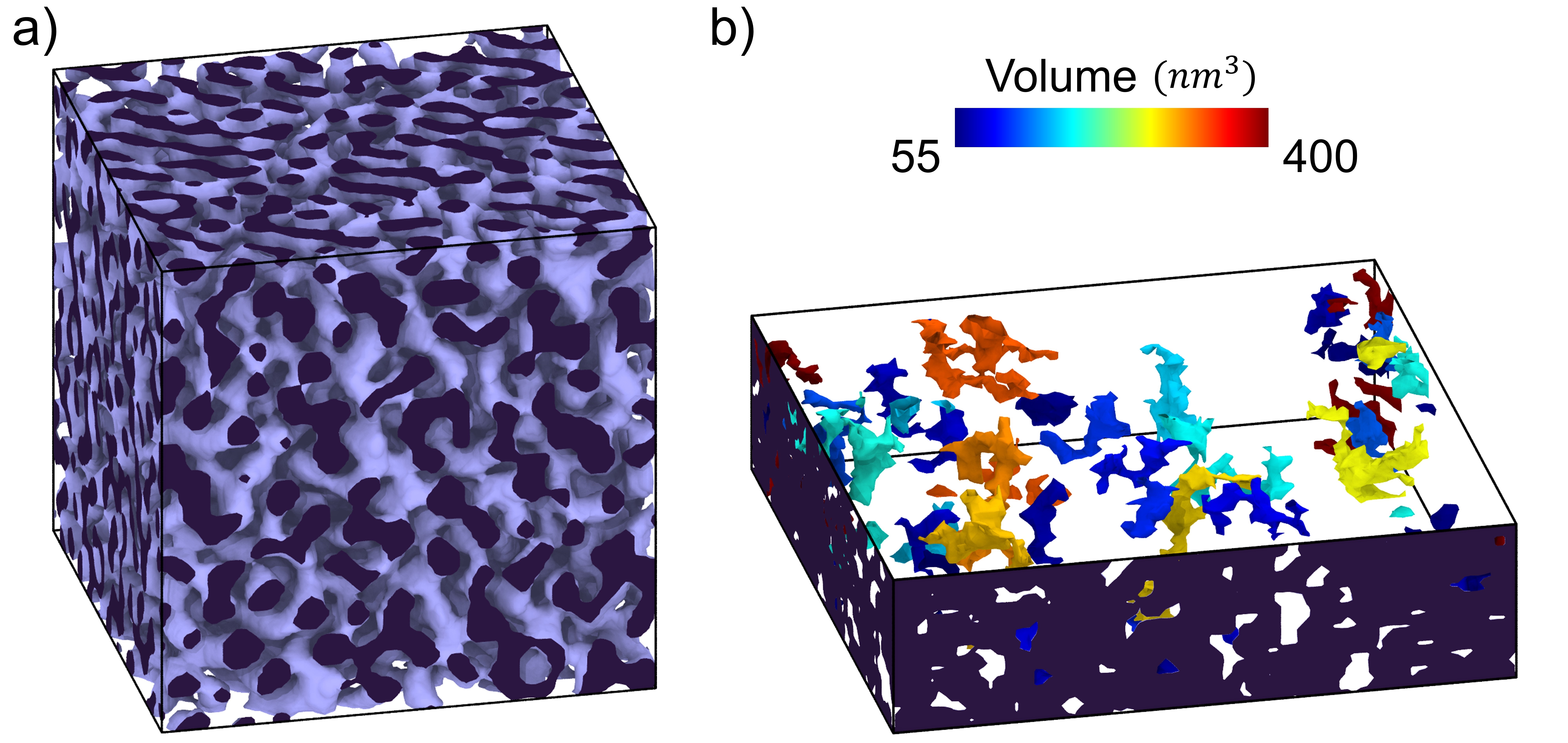}
    \caption{Reconstructed surface for the largest sample tested (250 $a_0$, $L=3.8~nm$). a) Undeformed state showing the free surfaces of the ligaments. b) At a strain of 0.7, the porous phase is no longer continuous. The sides of the box show a significant compaction of the foam, yet not fully densified. In fact, the interior of sample presents isolated voids in the material. The porous phase is now discontinuous and pores with different volume can be appreciated. In order to enhance the visibility, a number of interior surfaces were removed.}
    \label{fig:Percolation}
\end{figure}

For larger strains, the evolution of the genus shifts and starts decreasing. Notable differences in genus are also consistent with stress strain curves differentiation in the exponential densification regime, indicating variations in the densification rate. 
Note that the genus data is presented up to a strain of the order of 0.55. After that strain, the reconstructed surface is no longer closed, thus preventing the computation of genus over the whole sample. Analysis of the reconstructed surfaces reveals that, at such large strains, porosity is now distributed in the form of disconnected voids. 
The reconstructed surfaces we obtain at very high strains are indeed disconnected, as seen on Figure \ref{fig:Percolation}. 
This is correlated to a change of the foam from its original open-cell nature, to a partially closed-cell structure.


\section{Discussion}

The mechanical properties of nanoporous metals have often been studied using nanoporous gold as a prototypical example. Our results also show significant points of agreement with previous studies on that material. 
To begin with, the response clearly depicts three regimes, namely an elastic regime, a linear hardening regime, and an exponential regime, consistent with previous reports on np-Au \cite{saffarini2021ligament,jin2009deforming,ngo2015anomalous}. Deformation proceeds by ligament bending and plasticity takes place by dislocation nucleation at the ligament surfaces, in a dislocation accumulation scenario \cite{jin2009deforming,ngo2015anomalous,ruestes2016hardening,beets2019deformation,beets2018mechanical}. After yielding, some np Au works display a stress plateau \cite{farkas2013mechanical,ruestes2016hardening} whereas others display linear hardening \cite{ngo2015anomalous,saffarini2021ligament}. Our results show an agreement with the latter, probably due to the ligament size and relative density explored.

Mathesan and Mordehai \cite{mathesan2020size,mathesan2021yielding} used MD simulations to explore the influence of topology on the mechanical properties of np-Au nanopillars. In particular, they focused on the load-bearing network. They show a strong influence on the elastic properties \cite{mathesan2020size}. Later on, by detailed analysis of the genus of the yielded ligaments network, they conclude that the rate by which the strength increases with strain decreases gradually, in
correlation with a decrease in the genus of a subnetwork of the
unyielded ligaments \cite{mathesan2021yielding}. Although our study focuses on a different base material in the form of bulk np-W (in contrast to np-Au nanopillars in \cite{mathesan2020size,mathesan2021yielding}), a general agreement can be found. That is, \textit{(i)} after the onset of the plastic regime the genus increases until significant plastic strain has been reached; \textit{(ii)} a correlation of the genus with the stress strain response.

The linear hardening regime has been typically associated with dislocation activity as well as densification, whereas the exponential hardening is heavily influenced by latter. This has been shown in MD simulations of nanoporous gold under compression. 
Ngo et al. \cite{ngo2015anomalous} studies on np-Au provided compelling evidence pointing to dislocation storage as responsible for hardening behavior together with densification. Ruestes et al. \cite{ruestes2016hardening} and Saffarini et al. \cite{saffarini2021ligament} later presented more evidence and provided dislocation density-based equations relating dislocation density and solid fraction evolution. Here we provide an alternative interpretation for the different strain hardening regimes.

\begin{figure}[H]
    \centering
    \includegraphics[width=.75\textwidth]{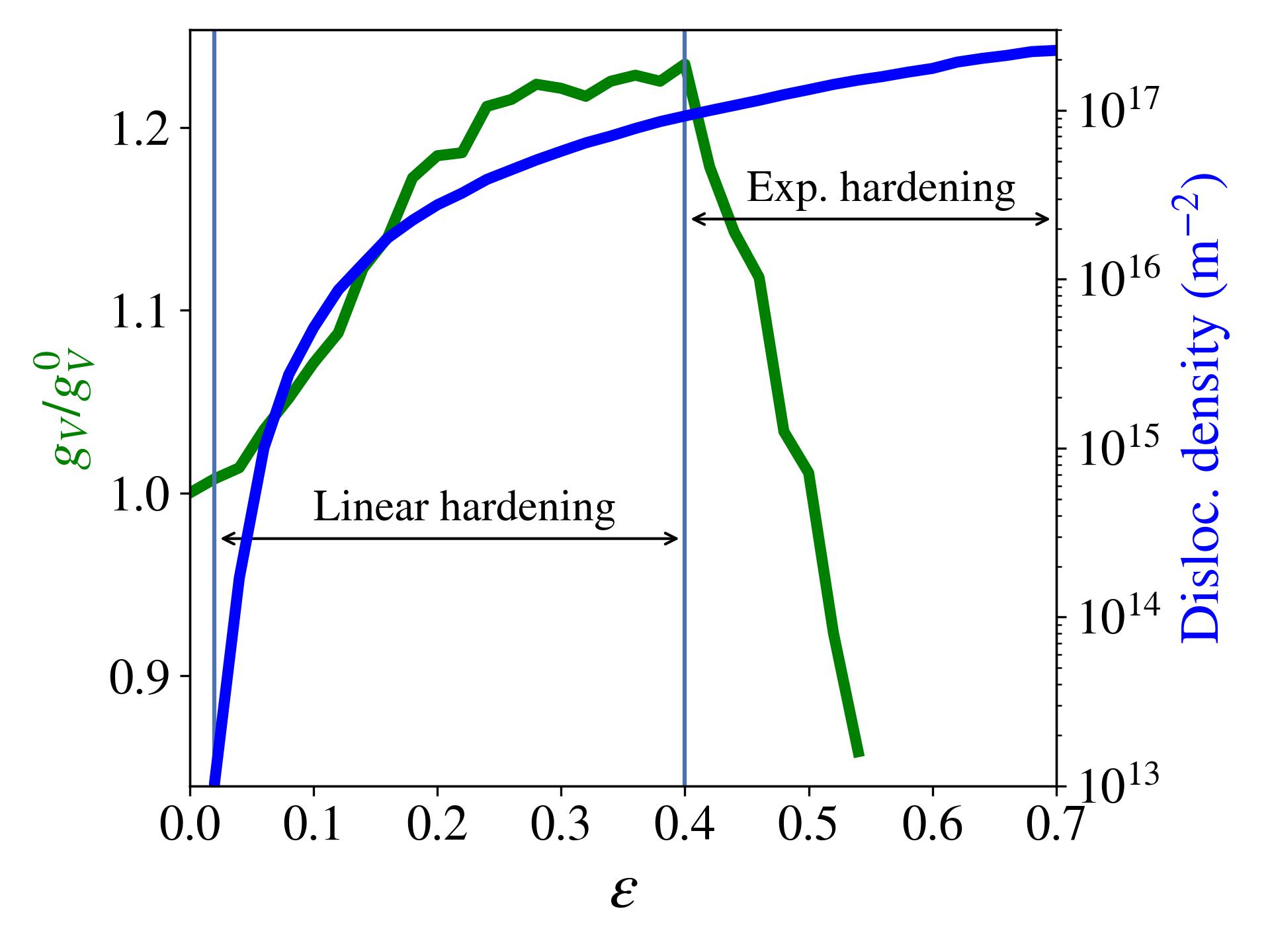}
    \caption{Comparative assessment of the evolution of the genus ratio and total dislocation density for the largest sample tested (250 $a_0$, $L=3.8~nm$) with respect to the different hardening regimes.}
    \label{fig:genus_vs_disden}
\end{figure}

Figure \ref{fig:genus_vs_disden} presents the evolution of the genus ration and the total dislocation density for the largest sample tested (250 $a_0$, $L=3.8~nm$). For interpretation purposes, the linear hardening regime and exponential hardening regime are indicated, as revealed by Fig. \ref{fig:stress_genus}. During the linear hardening regime, dislocation density rapidly increases, together with an increase in the genus ratio. The former indicates a dislocation-accumulation scenario, whereas the latter indicates a correlation of topology with linear hardening. During the exponential hardening regime, dislocation density keeps increasing in a less marked way due to dislocation densities approaching saturation limits ($\rho_D \sim 10^{17}/\text{m}{^{-2})}$. Interestingly, the genus ratio drops sharply during the exponential hardening regime. 
These results only allow for qualitative conclusions. Specific studies are needed to provide an adequate quantitative assessment of   the respective contributions of dislocations and topological changes for the two regimes.

With respect to refractory-based nanoporous metals, 
Xia and co-workers \cite{hu2022atomistic} used MD simulations to study the mechanical properties of nanoporous tungsten, another prototypical bcc metal with significant applications. By means of simulated tensile tests, they explored the effects of relative density, grain size , and temperature on the mechanical properties and
deformation mechanisms of np-W. They also report on dislocation mediated plasticity, as well as on a contribution of twinning as a complementary deformation mechanism. Similar deformation mechanisms were reported by Valencia and coworkers \cite{valencia2022nanoindentation} in their MD study of nanoporous W under nanoindentation. Our results are in agreement with these contributions.

\section{Conclusions}

In summary, we report on a computational study of the
compression behavior of nanoporous Ta using molecular dynamics simulations. Computational samples were constructed using our own implementation of a levelled-wave method. This software is released as an open-source tool, with multi-cpu and gpu capabilities, allowing for fast and efficient generation of nanoporous samples. Our main conclusions are: 
\begin{itemize}

    \item Consistent with previous studies on nanoporous gold, we find that the stress strain curves depict similar qualitative features. Namely, a non-linear elastic regime, early yielding, and linear hardening followed by an exponential hardening at large strains, associated with a high degree of densification.

    \item Plasticity takes place by dislocation activity in the expected bcc slip systems. Even though dislocations nucleate at the surface and most of them annihilate at a nearby surface after slip through the ligaments, deformation proceeds in a dislocation accumulation scenario. Ligament junctions provide enough volume for dislocation reaction, producing sessile dislocations, as well as dislocation storage. Densification of the sample favors dislocation retention. Twinning and vacancy formation was also found.

    \item The evolution of the genus was succesfully correlated to the stress strain behavior. Genus increases during the linear hardening regime, right after the onset of plasticity, indicating continuous densification of the sample. In contrast, the genus significantly drops at very large strains, that is during the exponential hardening regime. This is due to significant densification of the sample, shifting from an open cell foam to a closed cell foam.  

    \item Structural changes from open-cell foam to closed-cell foam are shown to be ligament-size dependent. Our analysis indicates that, whenever possible, tracking of topological invariants (here we use $g$) are effective measures to characterize the evolution of the sample and could provide valuable insights for the identification of deformation-induced changes. 

\end{itemize}


As a closing remark, our work suggests that fundamental understanding gained from np-Au studies is directly transferable to other nanoporous metals, albeit with differences in their deformation mechanisms. Most importantly, our analysis shows that topological metrics, genus being just one example, can provide fundamental insights into complex aspects of the deformation of nanoporous metals, such as the onset of densification and associated exponential hardening regime.


\section{Acknowledgements}

The work by NVvB was supported by an EVC scholarship from Consejo
Interuniversitario Nacional - Argentina. NVvB is currently supported by CONICET - Argentina. NVvB thanks F. Aquistapace for help on the use of FOAM EXPLORER. This work used the Toko Cluster from FCEN, UNCuyo, which is part of the SNCAD, MinCyT, Argentina. Support by ANPCyT PICT-2018-0773 is kindly acknowledged. The authors acknowledge valuable discussions with M.G. Del Pópolo, A. Lobos and E.M. Bringa during early stages of this manuscript.
This project has received funding from the European Union’s Horizon Europe research and innovation programme under the Marie Sklodowska-Curie grant agreement no. 101062254. Funded by the European Union.
Views and opinions expressed are however those of the author(s) only and
do not necessarily reflect those of the European Union. Neither the European
Union nor the granting authority can be held responsible for them.

\section{Data availability statement}

The computer code required to reproduce the microstructures presented here are available to download from \url{https://github.com/nicolasvazquez95/Nanofoams_Creator}. All other information is available from the authors upon a reasonable request.

\bibliographystyle{elsarticle-num}
\bibliography{references}

\begin{thebibliography}{10}
\expandafter\ifx\csname url\endcsname\relax
  \def\url#1{\texttt{#1}}\fi
\expandafter\ifx\csname urlprefix\endcsname\relax\def\urlprefix{URL }\fi
\expandafter\ifx\csname href\endcsname\relax
  \def\href#1#2{#2} \def\path#1{#1}\fi

\bibitem{ding2004metallic}
Y.~Ding, M.~Chen, J.~Erlebacher, Metallic mesoporous nanocomposites for
  electrocatalysis, Journal of the American Chemical Society 126~(22) (2004)
  6876--6877.

\bibitem{wittstock2023nanoporous}
G.~Wittstock, M.~B{\"a}umer, W.~Dononelli, T.~Kl{\"u}ner, L.~L{\"u}hrs,
  C.~Mahr, L.~V. Moskaleva, M.~Oezaslan, T.~Risse, A.~Rosenauer, et~al.,
  Nanoporous gold: From structure evolution to functional properties in
  catalysis and electrochemistry, Chemical Reviews 123~(10) (2023) 6716--6792.

\bibitem{biener2009surface}
J.~Biener, A.~Wittstock, L.~Zepeda-Ruiz, M.~Biener, V.~Zielasek, D.~Kramer,
  R.~Viswanath, J.~Weissm{\"u}ller, M.~B{\"a}umer, A.~Hamza,
  Surface-chemistry-driven actuation in nanoporous gold, Nature materials 8~(1)
  (2009) 47--51.

\bibitem{detsi2016metallic}
E.~Detsi, S.~H. Tolbert, S.~Punzhin, J.~T.~M. De~Hosson, Metallic muscles and
  beyond: nanofoams at work, Journal of materials science 51~(1) (2016)
  615--634.

\bibitem{bringa2011nanoporous}
E.~Bringa, J.~Monk, A.~Caro, A.~Misra, L.~Zepeda-Ruiz, M.~Duchaineau,
  F.~Abraham, M.~Nastasi, S.~Picraux, Y.~Wang, et~al., Are nanoporous materials
  radiation resistant?, Nano letters 12~(7) (2011) 3351--3355.

\bibitem{beyerlein2013radiation}
I.~Beyerlein, A.~Caro, M.~Demkowicz, N.~Mara, A.~Misra, B.~Uberuaga, Radiation
  damage tolerant nanomaterials, Materials today 16~(11) (2013) 443--449.

\bibitem{briot2019situ}
N.~J. Briot, M.~Kosmidou, R.~Dingreville, K.~Hattar, T.~J. Balk, In situ tem
  investigation of self-ion irradiation of nanoporous gold, Journal of
  Materials Science 54~(9) (2019) 7271--7287.

\bibitem{lionello2021mechanical}
D.~Lionello, J.~Ramallo, M.~Caro, Y.~Wang, C.~Sheehan, J.~Baldwin, J.~Nogan,
  A.~Caro, M.~Fuertes, C.~Ruestes, Mechanical properties of
  al2o3-functionalized nanoporous gold foams under irradiation, Journal of
  Materials Research (2021) 1--9.

\bibitem{biener2006size}
J.~Biener, A.~M. Hodge, J.~R. Hayes, C.~A. Volkert, L.~A. Zepeda-Ruiz, A.~V.
  Hamza, F.~F. Abraham, Size effects on the mechanical behavior of nanoporous
  au, Nano letters 6~(10) (2006) 2379--2382.

\bibitem{hodge2007scaling}
A.~Hodge, J.~Biener, J.~Hayes, P.~Bythrow, C.~Volkert, A.~Hamza, Scaling
  equation for yield strength of nanoporous open-cell foams, Acta Materialia
  55~(4) (2007) 1343--1349.

\bibitem{feng2009surface}
X.-Q. Feng, R.~Xia, X.~Li, B.~Li, Surface effects on the elastic modulus of
  nanoporous materials, Applied Physics Letters 94~(1) (2009) 011916.

\bibitem{sun2013mechanical}
X.-Y. Sun, G.-K. Xu, X.~Li, X.-Q. Feng, H.~Gao, Mechanical properties and
  scaling laws of nanoporous gold, Journal of Applied Physics 113~(2) (2013)
  023505.

\bibitem{farkas2013mechanical}
D.~Farkas, A.~Caro, E.~Bringa, D.~Crowson, Mechanical response of nanoporous
  gold, Acta Materialia 61~(9) (2013) 3249--3256.

\bibitem{briot2014mechanical}
N.~J. Briot, T.~Kennerknecht, C.~Eberl, T.~J. Balk, Mechanical properties of
  bulk single crystalline nanoporous gold investigated by millimetre-scale
  tension and compression testing, Philosophical Magazine 94~(8) (2014)
  847--866.

\bibitem{luhrs2016elastic}
L.~L{\"u}hrs, C.~Soyarslan, J.~Markmann, S.~Bargmann, J.~Weissm{\"u}ller,
  Elastic and plastic poisson’s ratios of nanoporous gold, Scripta Materialia
  110 (2016) 65--69.

\bibitem{ruestes2016hardening}
C.~J. Ruestes, D.~Farkas, A.~Caro, E.~M. Bringa, Hardening under compression in
  au foams, Acta Materialia 108 (2016) 1--7.

\bibitem{jiao2017deformation}
J.~Jiao, N.~Huber, Deformation mechanisms in nanoporous metals: Effect of
  ligament shape and disorder, Computational Materials Science 127 (2017)
  194--203.

\bibitem{badwe2017mechanical}
N.~Badwe, X.~Chen, K.~Sieradzki, Mechanical properties of nanoporous gold in
  tension, Acta Materialia 129 (2017) 251--258.

\bibitem{luhrs2017plastic}
L.~L{\"u}hrs, B.~Zandersons, N.~Huber, J.~Weissm{\"u}ller, Plastic poisson’s
  ratio of nanoporous metals: a macroscopic signature of tension--compression
  asymmetry at the nanoscale, Nano letters 17~(10) (2017) 6258--6266.

\bibitem{burckert2017uniaxial}
M.~B{\"u}rckert, N.~J. Briot, T.~J. Balk, Uniaxial compression testing of bulk
  nanoporous gold, Philosophical Magazine 97~(15) (2017) 1157--1178.

\bibitem{pia2018nanoporous}
G.~Pia, M.~Carta, F.~Delogu, Nanoporous au foams: Variation of effective
  young's modulus with ligament size, Scripta Materialia 144 (2018) 22--26.

\bibitem{beets2019deformation}
N.~Beets, D.~Farkas, S.~Corcoran, Deformation mechanisms and scaling relations
  in the mechanical response of nano-porous au, Acta Materialia 165 (2019)
  626--637.

\bibitem{wilkerson2019anomalous}
J.~W. Wilkerson, Anomalous size effects in nanoporous materials induced by high
  surface energies, Journal of Materials Research 34~(13) (2019) 2337--2346.

\bibitem{jin2018mechanical}
H.-J. Jin, J.~Weissm{\"u}ller, D.~Farkas, Mechanical response of nanoporous
  metals: A story of size, surface stress, and severed struts, Mrs Bulletin
  43~(1) (2018) 35--42.

\bibitem{biener2011ald}
M.~M. Biener, J.~Biener, A.~Wichmann, A.~Wittstock, T.~F. Baumann,
  M.~B{\"a}umer, A.~V. Hamza, Ald functionalized nanoporous gold: thermal
  stability, mechanical properties, and catalytic activity, Nano letters 11~(8)
  (2011) 3085--3090.

\bibitem{li2019topology}
Y.~Li, B.-N.~D. Ng{\^o}, J.~Markmann, J.~Weissm{\"u}ller, Topology evolution
  during coarsening of nanoscale metal network structures, Physical review
  materials 3~(7) (2019) 076001.

\bibitem{mccue2015frontiers}
I.~D. McCue, Frontiers of dealloying-novel processing for advanced materials,
  Ph.D. thesis, Johns Hopkins University (2015).

\bibitem{geslin2015topology}
P.-A. Geslin, I.~McCue, B.~Gaskey, J.~Erlebacher, A.~Karma, Topology-generating
  interfacial pattern formation during liquid metal dealloying, Nature
  communications 6~(1) (2015) 8887.

\bibitem{mccue2016kinetics}
I.~McCue, B.~Gaskey, P.-A. Geslin, A.~Karma, J.~Erlebacher, Kinetics and
  morphological evolution of liquid metal dealloying, Acta Materialia 115
  (2016) 10--23.

\bibitem{hou2018nanoporous}
C.~Hou, J.~Wang, H.~Wang, X.~Liu, S.~Liang, X.~Song, Z.~Nie, Nanoporous
  tungsten with tailorable microstructure and high thermal stability,
  International Journal of Refractory Metals and Hard Materials 77 (2018)
  128--131.

\bibitem{kosmidou2019vacuum}
M.~Kosmidou, M.~J. Detisch, T.~L. Maxwell, T.~J. Balk, Vacuum thermal
  dealloying of magnesium-based alloys for fabrication of nanoporous refractory
  metals, MRS Communications 9~(1) (2019) 144--149.

\bibitem{gaskey2019self}
B.~Gaskey, I.~McCue, A.~Chuang, J.~Erlebacher, Self-assembled porous
  metal-intermetallic nanocomposites via liquid metal dealloying, Acta
  Materialia 164 (2019) 293--300.

\bibitem{chuang2022powder}
A.~Chuang, J.~Baris, C.~Ott, I.~McCue, J.~Erlebacher, A powder metallurgy
  approach to liquid metal dealloying with applications in additive
  manufacturing, Acta Materialia 238 (2022) 118213.

\bibitem{zhao2020tailoring}
M.~Zhao, I.~Issa, M.~J. Pfeifenberger, M.~Wurmshuber, D.~Kiener, Tailoring
  ultra-strong nanocrystalline tungsten nanofoams by reverse phase dissolution,
  Acta Materialia 182 (2020) 215--225.

\bibitem{zhao2021open}
M.~Zhao, K.~Schlueter, M.~Wurmshuber, M.~Reitgruber, D.~Kiener, Open-cell
  tungsten nanofoams: Scaling behavior and structural disorder dependence of
  young’s modulus and flow strength, Materials \& Design 197 (2021) 109187.

\bibitem{detsi2011specific}
E.~Detsi, E.~De~Jong, A.~Zinchenko, Z.~Vukovi{\'c}, I.~Vukovi{\'c}, S.~Punzhin,
  K.~Loos, G.~Ten~Brinke, H.~De~Raedt, P.~Onck, et~al., On the specific surface
  area of nanoporous materials, Acta Materialia 59~(20) (2011) 7488--7497.

\bibitem{liu2016interpreting}
L.-Z. Liu, X.-L. Ye, H.-J. Jin, Interpreting anomalous low-strength and
  low-stiffness of nanoporous gold: Quantification of network connectivity,
  Acta Materialia 118 (2016) 77--87.

\bibitem{mangipudi2016topology}
K.~Mangipudi, E.~Epler, C.~Volkert, Topology-dependent scaling laws for the
  stiffness and strength of nanoporous gold, Acta Materialia 119 (2016)
  115--122.

\bibitem{lilleodden2018topological}
E.~T. Lilleodden, P.~W. Voorhees, On the topological, morphological, and
  microstructural characterization of nanoporous metals, MRS Bulletin 43~(1)
  (2018) 20--26.

\bibitem{li2018mechanical}
J.~Li, Y.~Xian, H.~Zhou, R.~Wu, G.~Hu, R.~Xia, Mechanical properties of
  nanocrystalline nanoporous gold complicated by variation of grain and
  ligament: A molecular dynamics simulation, Science China Technological
  Sciences 61 (2018) 1353--1363.

\bibitem{saffarini2021ligament}
M.~H. Saffarini, G.~Z. Voyiadjis, C.~J. Ruestes, M.~Yaghoobi, Ligament size
  dependency of strain hardening and ductility in nanoporous gold,
  Computational Materials Science 186 (2021) 109920.

\bibitem{yildiz2020strain}
Y.~O. Yildiz, A.~Ahadi, M.~Kirca, Strain rate effects on tensile and
  compression behavior of nano-crystalline nanoporous gold: A molecular dynamic
  study, Mechanics of Materials 143 (2020) 103338.

\bibitem{voyiadjis2021characterization}
G.~Z. Voyiadjis, M.~H. Saffarini, C.~J. Ruestes, Characterization of the strain
  rate effect under uniaxial loading for nanoporous gold, Computational
  Materials Science 194 (2021) 110425.

\bibitem{li2020nanoindentation}
J.~Li, B.~Lu, Y.~Zhang, H.~Zhou, G.~Hu, R.~Xia, Nanoindentation response of
  nanocrystalline copper via molecular dynamics: Grain-size effect, Materials
  Chemistry and Physics 241 (2020) 122391.

\bibitem{saffarini2021temperature}
M.~H. Saffarini, G.~Z. Voyiadjis, C.~J. Ruestes, Temperature effect on
  nanoporous gold under uniaxial tension and compression, Computational
  Materials Science 200 (2021) 110766.

\bibitem{ngo2015anomalous}
B.-N.~D. Ng{\^o}, A.~Stukowski, N.~Mameka, J.~Markmann, K.~Albe,
  J.~Weissm{\"u}ller, Anomalous compliance and early yielding of nanoporous
  gold, Acta Materialia 93 (2015) 144--155.

\bibitem{beets2018mechanical}
N.~Beets, D.~Farkas, Mechanical response of au foams of varying porosity from
  atomistic simulations, JOM (2018) 1--7.

\bibitem{mathesan2020size}
S.~Mathesan, D.~Mordehai, Size-dependent elastic modulus of nanoporous au
  nanopillars, Acta Materialia 185 (2020) 441--452.

\bibitem{mathesan2021yielding}
S.~Mathesan, D.~Mordehai, On the yielding and densification of nanoporous au
  nanopillars in molecular dynamics simulations, Computational Materials
  Science 191 (2021) 110307.

\bibitem{saffarini2021scaling}
M.~H. Saffarini, G.~Z. Voyiadjis, C.~J. Ruestes, Scaling laws for nanoporous
  metals under uniaxial loading, Journal of Materials Research 36~(13) (2021)
  2729--2741.

\bibitem{beets2020fracture}
N.~Beets, J.~Stuckner, M.~Murayama, D.~Farkas, Fracture in nanoporous gold: An
  integrated computational and experimental study, Acta Materialia 185 (2020)
  257--270.

\bibitem{he2022mechanical}
L.~He, N.~Abdolrahim, Mechanical enhancement of graded nanoporous structure,
  Journal of Engineering Materials and Technology 144~(1) (2022).

\bibitem{shi2021scaling}
S.~Shi, Y.~Li, B.-N. Ngo-Dinh, J.~Markmann, J.~Weissm{\"u}ller, Scaling
  behavior of stiffness and strength of hierarchical network nanomaterials,
  Science 371~(6533) (2021) 1026--1033.

\bibitem{hodge2006characterization}
A.~M. Hodge, J.~R. Hayes, J.~A. Caro, J.~Biener, A.~V. Hamza, Characterization
  and mechanical behavior of nanoporous gold, Advanced Engineering Materials
  8~(9) (2006) 853.

\bibitem{newman1999alloy}
R.~Newman, S.~Corcoran, J.~Erlebacher, M.~Aziz, K.~Sieradzki, Alloy corrosion,
  Mrs Bulletin 24~(7) (1999) 24--28.

\bibitem{crowson2007geometric}
D.~A. Crowson, D.~Farkas, S.~G. Corcoran, Geometric relaxation of nanoporous
  metals: The role of surface relaxation, Scripta materialia 56~(11) (2007)
  919--922.

\bibitem{crowson2009mechanical}
D.~A. Crowson, D.~Farkas, S.~G. Corcoran, Mechanical stability of nanoporous
  metals with small ligament sizes, Scripta Materialia 61~(5) (2009) 497--499.

\bibitem{soyarslan20183d}
C.~Soyarslan, S.~Bargmann, M.~Pradas, J.~Weissm{\"u}ller, 3d stochastic
  bicontinuous microstructures: Generation, topology and elasticity, Acta
  materialia 149 (2018) 326--340.

\bibitem{liu2019efficient}
C.~Liu, P.~S. Branicio, Efficient generation of non-cubic stochastic periodic
  bicontinuous nanoporous structures, Computational Materials Science 169
  (2019) 109101.

\bibitem{guillotte2019fully}
M.~Guillotte, J.~Godet, L.~Pizzagalli, A fully molecular dynamics-based method
  for modeling nanoporous gold, Computational Materials Science 161 (2019)
  135--142.

\bibitem{cahn1965phase}
J.~W. Cahn, Phase separation by spinodal decomposition in isotropic systems,
  The Journal of Chemical Physics 42~(1) (1965) 93--99.

\bibitem{aparicio2020foamexplorer}
E.~Aparicio, E.~N. Mill{\'a}n, C.~J. Ruestes, E.~M. Bringa, Foamexplorer:
  Automated measurement of ligaments and voids for atomistic systems,
  Computational Materials Science 185 (2020) 109942.

\bibitem{plimpton1995fast}
S.~Plimpton, Fast parallel algorithms for short-range molecular dynamics,
  Journal of computational physics 117~(1) (1995) 1--19.

\bibitem{dai2006extended}
X.~Dai, Y.~Kong, J.~Li, B.~Liu, Extended finnis--sinclair potential for bcc and
  fcc metals and alloys, Journal of Physics: Condensed Matter 18~(19) (2006)
  4527.

\bibitem{tang2011growth}
Y.~Tang, E.~M. Bringa, B.~A. Remington, M.~A. Meyers, Growth and collapse of
  nanovoids in tantalum monocrystals, Acta Materialia 59~(4) (2011) 1354--1372.

\bibitem{tang2012growth}
Y.~Tang, E.~M. Bringa, B.~Remington, M.~Meyers, Growth and collapse of
  nanovoids in tantalum monocrystals loaded at high strain rate, in: AIP
  Conference Proceedings, Vol. 1426, American Institute of Physics, 2012, pp.
  1255--1258.

\bibitem{ruestes2014atomistic}
C.~J. Ruestes, A.~Stukowski, Y.~Tang, D.~Tramontina, P.~Erhart, B.~Remington,
  H.~Urbassek, M.~A. Meyers, E.~M. Bringa, Atomistic simulation of tantalum
  nanoindentation: Effects of indenter diameter, penetration velocity, and
  interatomic potentials on defect mechanisms and evolution, Materials Science
  and Engineering: A 613 (2014) 390--403.

\bibitem{ruestes2013atomistic}
C.~J. Ruestes, E.~M. Bringa, A.~Stukowski, J.~R. Nieva, G.~Bertolino, Y.~Tang,
  M.~Meyers, Atomistic simulation of the mechanical response of a nanoporous
  body-centered cubic metal, Scripta Materialia 68~(10) (2013) 817--820.

\bibitem{tramontina2014orientation}
D.~Tramontina, C.~Ruestes, Y.~Tang, E.~Bringa, Orientation-dependent response
  of defective tantalum single crystals, Computational Materials Science 90
  (2014) 82--88.

\bibitem{dou2011deformation}
R.~Dou, B.~Derby, Deformation mechanisms in gold nanowires and nanoporous gold,
  Philosophical Magazine 91~(7-9) (2011) 1070--1083.

\bibitem{stukowski2010visualization}
A.~Stukowski, Visualization and analysis of atomistic simulation data with
  ovito--the open visualization tool, Modelling and Simulation in Materials
  Science and Engineering 18~(1) (2010) 015012.

\bibitem{stukowski2012automated}
A.~Stukowski, V.~V. Bulatov, A.~Arsenlis, Automated identification and indexing
  of dislocations in crystal interfaces, Modelling and Simulation in Materials
  Science and Engineering 20~(8) (2012) 085007.

\bibitem{gibson1997cellular}
L.~J. Gibson, M.~F. Ashby, Cellular solids: structure and properties, Cambridge
  university press, 1997.

\bibitem{jin2009deforming}
H.-J. Jin, L.~Kurmanaeva, J.~Schmauch, H.~R{\"o}sner, Y.~Ivanisenko,
  J.~Weissm{\"u}ller, Deforming nanoporous metal: Role of lattice coherency,
  Acta Materialia 57~(9) (2009) 2665--2672.

\bibitem{mathesan2023displacive}
S.~Mathesan, D.~Mordehai, Displacive-diffusive plasticity in nanoporous gold
  nanowires under tensile creep, Scripta Materialia 224 (2023) 115106.

\bibitem{soltani2018mechanism}
S.~Soltani, N.~Abdolrahim, P.~Sepehrband, Mechanism of intrinsic diffusion in
  the core of screw dislocations in fcc metals--a molecular dynamics study,
  Computational Materials Science 144 (2018) 50--55.

\bibitem{dupraz2018dislocation}
M.~Dupraz, Z.~Sun, C.~Brandl, H.~Van~Swygenhoven, Dislocation interactions at
  reduced strain rates in atomistic simulations of nanocrystalline al, Acta
  Materialia 144 (2018) 68--79.

\bibitem{murr1997shock}
L.~Murr, M.~Meyers, C.-S. Niou, Y.~Chen, S.~Pappu, C.~Kennedy, Shock-induced
  deformation twinning in tantalum, Acta materialia 45~(1) (1997) 157--175.

\bibitem{stukowski2014computational}
A.~Stukowski, Computational analysis methods in atomistic modeling of crystals,
  Jom 66~(3) (2014) 399--407.

\bibitem{krone2012fast}
M.~Krone, J.~E. Stone, T.~Ertl, K.~Schulten, Fast visualization of gaussian
  density surfaces for molecular dynamics and particle system trajectories.,
  EuroVis (Short Papers) 10 (2012) 067--071.

\bibitem{mangipudi2016fib}
K.~R. Mangipudi, V.~Radisch, L.~Holzer, C.~A. Volkert, A fib-nanotomography
  method for accurate 3d reconstruction of open nanoporous structures,
  Ultramicroscopy 163 (2016) 38--47.

\bibitem{hu2022atomistic}
Y.~Hu, J.~Xu, L.~Su, Y.~Zhang, S.~Ding, R.~Xia, Atomistic simulations of
  mechanical characteristics dependency on relative density, grain size, and
  temperature of nanoporous tungsten, Physica Scripta 98~(1) (2022) 015715.

\bibitem{valencia2022nanoindentation}
F.~J. Valencia, R.~Ortega, R.~I. Gonzalez, E.~M. Bringa, M.~Kiwi, C.~J.
  Ruestes, Nanoindentation of nanoporous tungsten: A molecular dynamics
  approach, Computational Materials Science 209 (2022) 111336.

\end{thebibliography}

\section{Appendix}\label{Appendix}

\subsection{Computational performance and microstructures}

\begin{figure}[h]
    \centering
    \includegraphics[width=0.8\textwidth]{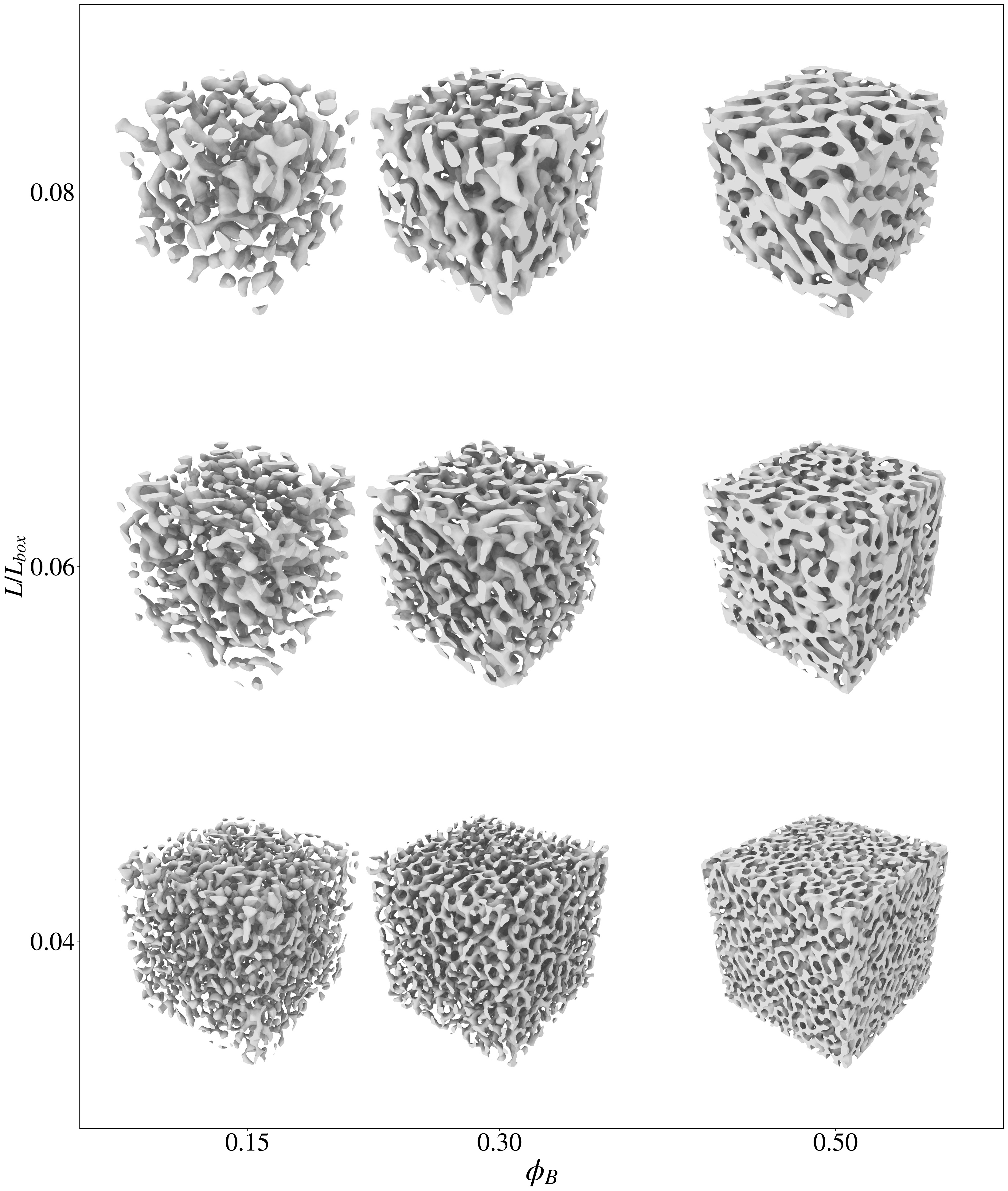}
    \caption{Microstructures generated with our open-source implementation of the levelled-wave method \cite{soyarslan20183d}. Note the percolation of the solid phase on the left hand-side column.}
    \label{fig:panel_9espumas}
\end{figure}

Figure  \ref{fig:panel_9espumas} displays a limited set of microstructures generated by our implementation of the levelled-wave method \cite{soyarslan20183d}.  For a small solid fraction $\phi_B$, or for ligament sizes within the order of magnitude of the lattice parameter, it is possible to end up with a microstructure that it is below the percolation limit. In other words, the solid phase is no longer continuous \cite{soyarslan20183d}. For solid fractions above 0.2 and for  $L/L_{\text{box}}$ above 0.04 ($L_{\text{box}}$ 
 being the simulation box size), our implementation produces stochastic and fully bicontinuous microstructures. 

\begin{figure}[h]
    \centering
    \includegraphics[width=.7\textwidth]{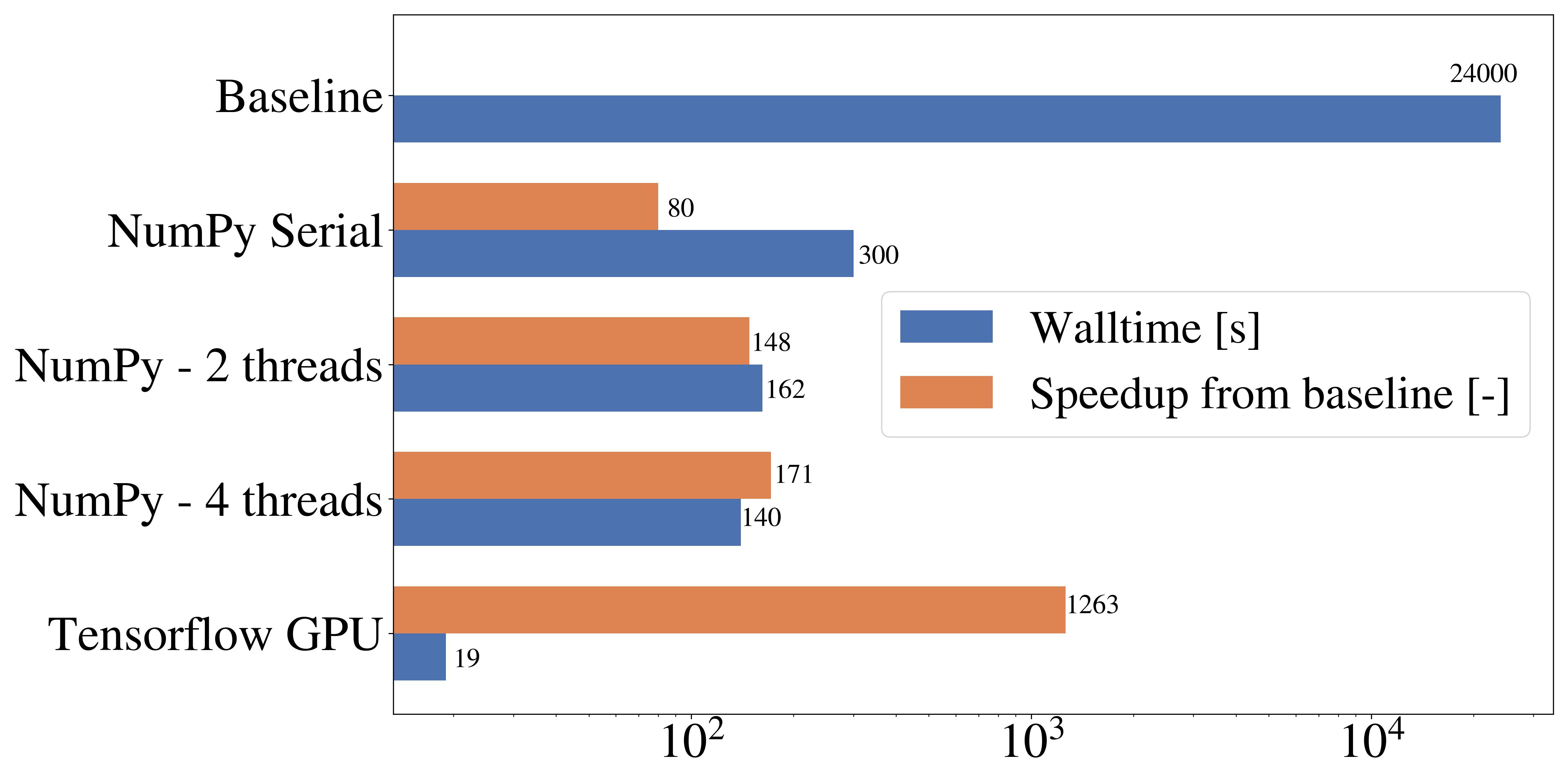}
    \caption{Benchmarking results using four different code versions. Numpy/CPU Tests were performed on an AMD Ryzen3 4000-series laptop. GPU tests were performed on a NVidia K80 GPU (4.1 TFlops).}
    \label{fig:benchmark1}
\end{figure}

To check the computational efficiency of our implementation, tests were performed using 4 different implementations. Figure  \ref{fig:benchmark1} presents the results of the tests performed for the same foam input parameters ($100^3$ simple cubic cells, 48 directions, 10000 waves) Blue bars indicate the execution  \textit{``Walltime''} in seconds, while orange bars indicate the  \textit{speedup}. This is a metric used to measure the scalability of a computational code, defined as $\operatorname{speedup} = \frac{t_1}{t_N}$. $t_1$ corresponds to the time required to run the code on a single processor, $t_N$ is the time required on $N$ processors. 
The results show a clear acceleration after the multi-CPU implementation. By recasting nested-loops into matrix operations, a GPU implementation is straigthforward by means of TensorFlow packages. This last implementation displays a notorious speedup.  




\begin{figure}[h]
    \centering
    \includegraphics[width=\textwidth]{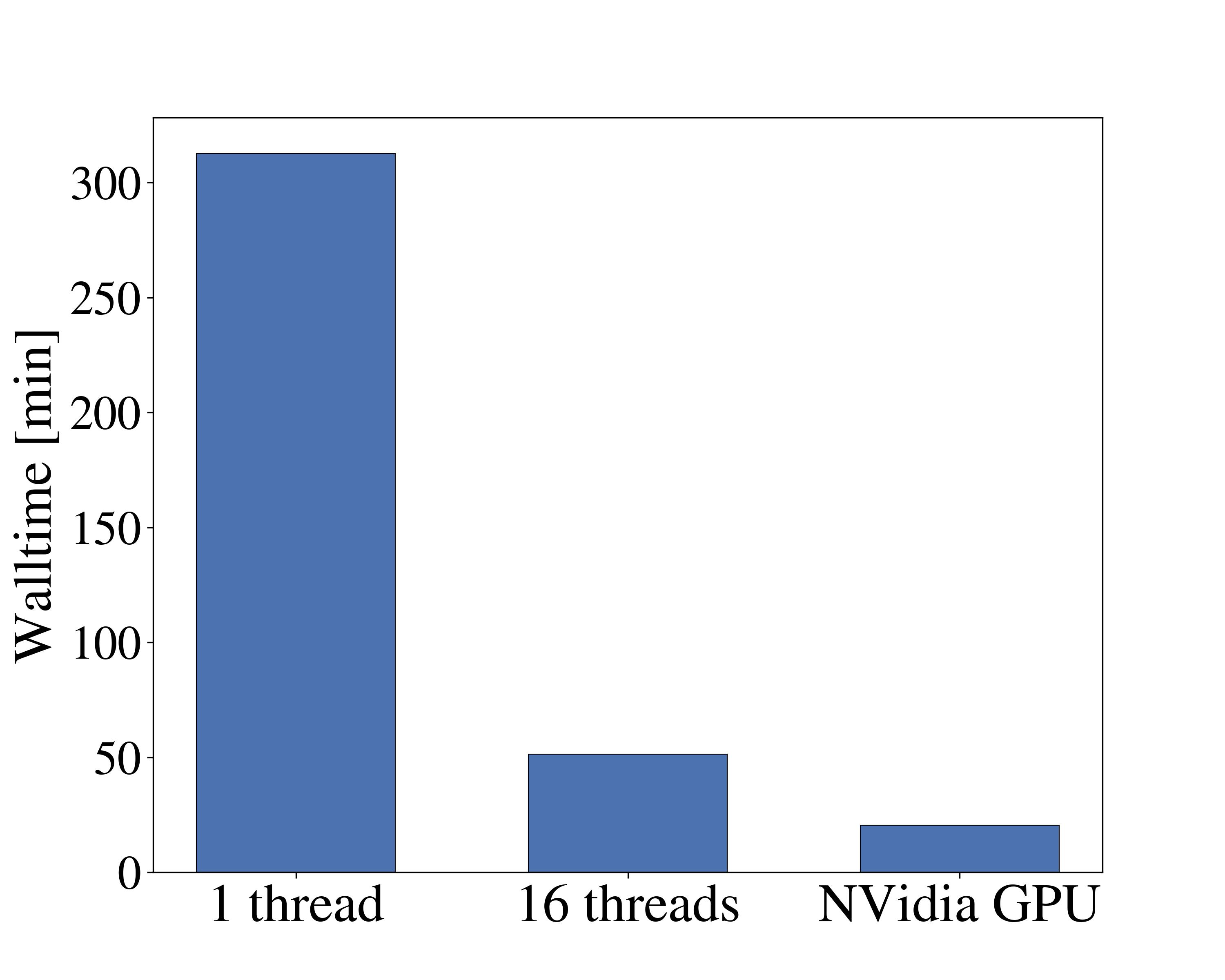}
    \caption{\textbf{(a)} Comparison between three generation runs executed on a single core (serial), parallely on 16 cores (multi-CPU) and on GPU. CPU characteristics: 8 cores / 16 threads, AMD Ryzen 7 processor, 64 GB RAM). GPU: NVidia K80 through GoogleColab.}
    \label{fig:Benchmark2}
\end{figure}

With the final CPU and GPU implementations, a last comparison of execution times was performed. This was done on 1 core and 16 cores for the CPU implementation. Tests were performed on TOKO FCEN UNCuyo computer cluster (1 node, AMD Ryzen 7, 64 GB Ram, 8 cores / 16 threads). The sample contained $300^3$ unit cells and 20000 were used. The comparison includes walltime for a test on a K80 NVidia GPU, used over GoogleColab. Excellent performance was obtained (See Fig. \ref{fig:Benchmark2}). 

Finally, our tests made use of 4GB of RAM (max) for our 16 threads CPU tests. The GPU implementation made a peak memory usage of 16 GB on the GPU plus an additional 3GB of RAM.
\end{document}